\documentclass[12pt]{article}

\usepackage{amssymb}
\include{epsf}

\newcommand{\newsection}{\setcounter{equation}{0}\section}

\def\appendix#1{\addtocounter{section}{1}\setcounter{equation}{0}
\renewcommand{\thesection}{\Alph{section}}
\section*{Appendix\thesection\protect\indent \parbox[t]{11.715cm} {#1}}
\addcontentsline{toc}{section}{Appendix \thesection\ \ \ #1} }
\newcommand{\complex}{{\bb C}} 
\newcommand{\complexs}{{\bbs C}} 
\newcommand{\zed}{{\bb Z}} 
\newcommand{\unitary}{{\bb U}} 
\newcommand{\real}{{\bb R}} 
\newcommand{\mat}{{\bb M}} 
\newcommand{\id}{{1\!\!1}} 
\def\alg{{\cal A}}
\def\hil{{\rm Gr}}

\font\mybb=msbm10 at 12pt
\def\bb#1{\hbox{\mybb#1}}
\font\mybbs=msbm10 at 9pt
\def\bbs#1{\hbox{\mybbs#1}}

\def\nn{\nonumber}
\newcommand{\tr}[1]{\:{\rm tr}\,#1}
\newcommand{\Tr}[1]{\:{\rm Tr}\,#1}
\def\e{{\,\rm e}\,}

\def\be{\begin{equation}}
\def\ee{\end{equation}}
\def\bea{\begin{eqnarray}}
\def\eea{\end{eqnarray}}
\def\bd{\begin{displaymath}}
\def\ed{\end{displaymath}}

\def\DD{{\rm D}}
\def\dd{{\rm d}}
\def\CC{{\sf C}}
\def\QQ{{\rm Q}}
\def\ii{{\,{\rm i}\,}}

\def\K{{{\rm K}_0}}
\def\K1{{{\rm K}_1}}
\def\bfp{{\vec p\,}}
\def\bfq{{\vec q\,}}
\def\bfN{{\vec N}}

\makeatletter
\newdimen\normalarrayskip              
\newdimen\minarrayskip                 
\normalarrayskip\baselineskip
\minarrayskip\jot
\newif\ifold             \oldtrue            \def\new{\oldfalse}
\def\arraymode{\ifold\relax\else\displaystyle\fi} 
\def\@arrayskip{\ifold\baselineskip\z@\lineskip\z@
     \else
     \baselineskip\minarrayskip\lineskip2\minarrayskip\fi}
\def\@arrayclassz{\ifcase \@lastchclass \@acolampacol \or
\@ampacol \or \or \or \@addamp \or
   \@acolampacol \or \@firstampfalse \@acol \fi
\edef\@preamble{\@preamble
  \ifcase \@chnum
     \hfil$\relax\arraymode\@sharp$\hfil
     \or $\relax\arraymode\@sharp$\hfil
     \or \hfil$\relax\arraymode\@sharp$\fi}}
\def\@array[#1]#2{\setbox\@arstrutbox=\hbox{\vrule
     height\arraystretch \ht\strutbox
     depth\arraystretch \dp\strutbox
     width\z@}\@mkpream{#2}\edef\@preamble{\halign \noexpand\@halignto
\bgroup \tabskip\z@ \@arstrut \@preamble \tabskip\z@ \cr}%
\let\@startpbox\@@startpbox \let\@endpbox\@@endpbox
  \if #1t\vtop \else \if#1b\vbox \else \vcenter \fi\fi
  \bgroup \let\par\relax
  \let\@sharp##\let\protect\relax
  \@arrayskip\@preamble}
\makeatother

\newcommand{\beq}{\begin{eqnarray}}
\newcommand{\eeq}{\end{eqnarray}}

\newcommand{\g}{\gamma}

\begin{document}
\begin{titlepage}
\begin{flushright}

\baselineskip=12pt
MCTP--02--17 \\ HWM--02--12\\ EMPG--02--05\\
\hfill{ }\\ March 2002
\end{flushright}

\begin{center}

\baselineskip=24pt

{\Large\bf Instanton Expansion of\\ Noncommutative Gauge Theory\\ in
Two Dimensions}

\baselineskip=14pt

\vspace{1cm}

{\bf L.D. Paniak}
\\[3mm]
{\it Michigan Center for Theoretical Physics\\ University of
Michigan\\ Ann Arbor, Michigan 48109-1120, U.S.A.}\\  {\tt
paniak@umich.edu}
\\[6mm]

{\bf R.J. Szabo}
\\[3mm]
{\it Department of Mathematics\\ Heriot-Watt University\\ Riccarton,
Edinburgh EH14 4AS, U.K.}\\  {\tt R.J.Szabo@ma.hw.ac.uk}
\\[30mm]

\end{center}

\begin{abstract}

\baselineskip=12pt

We show that noncommutative gauge theory in two dimensions
is an exactly solvable model. A cohomological formulation of gauge
theory defined on the noncommutative torus is used to show that its
quantum partition function can be written as a sum over contributions from
classical solutions. We derive an explicit formula for the
partition function of Yang-Mills theory defined on a projective module
for arbitrary noncommutativity parameter $\theta$ which is
manifestly invariant under gauge Morita equivalence.
The energy observables are shown to be smooth functions of $\theta$.
The construction of noncommutative instanton contributions to the path
integral is described in some detail. In general, there are infinitely
many gauge inequivalent contributions of fixed topological charge,
along with a finite number of
quantum fluctuations about each instanton. The associated moduli spaces are
combinations of symmetric products of an ordinary two-torus whose
orbifold singularities are not resolved by noncommutativity. In
particular, the weak coupling limit of the gauge theory is independent
of $\theta$ and computes the symplectic volume of the moduli space of constant
curvature connections on the noncommutative torus.

\end{abstract}

\end{titlepage}

{\baselineskip=12pt
\tableofcontents}

\newpage

\newsection{Introduction and Summary}

Quantum field theories on noncommutative spacetimes provide field
theoretical contexts in which to study the dynamics of D-branes, while
at the same time retaining the non-locality inherent in string theory
(see~\cite{ks}--\cite{sz1} for reviews). Recent studies of these field
theories have raised many questions regarding their existence and
properties, and even after extensive study there remain numerous
questions concerning the new phenomena they exhibit even in the
simplest cases. Of particular interest is Yang-Mills theory defined on
a noncommutative torus which serves as an effective description of
open strings propagating in flat backgrounds. In particular,
noncommutative gauge theory on a two-dimensional torus describes
codimension two vortex bound states of D-branes inside D-branes. In
this paper we will show that this quantum field theory is exactly
solvable and explicitly evaluate its partition function. Various
non-trivial aspects of noncommutative gauge theories in two dimensions
may be found in~\cite{poly}--\cite{gsv}.

The commutative version of this theory has a well-known history as an
exactly solvable model, which gives the first example of a confining
gauge theory whose infrared limit can be reformulated analytically as
a string theory (see~\cite{cmr,abdalla} for reviews). The key
feature of two dimensions is that there are no gluons and the theory
must be investigated on spacetimes of non-trivial topology or with
Wilson loops in order to see any degrees of freedom. This suppression
of degrees of freedom owes to the fact that the group of local
symmetries of two-dimensional Yang-Mills theory contains not only
local gauge invariance, but also invariance under area-preserving
diffeomorphisms. Of the several different methods for solving this
quantum field theory, a particularly fruitful approach is provided by
the lattice formulation~\cite{wilson}. Using the area-preserving
diffeomorphism invariance, the heat kernel expansion of the disk
amplitude may be interpreted as a wavefunction for a plaquette. The
fusion rules for group characters allow one to glue together
disconnected plaquettes. The basic plaquette Boltzmann weight in this
way turns out to be renormalization group invariant~\cite{witten1}, so
that the lattice gauge theory reproduces exactly the continuum answer.

While a lattice formulation of noncommutative Yang-Mills theory does
exist~\cite{amns}, it does not exhibit an obvious self-similarity
property as its commutative counterpart does. The non-locality of the
star-product mixes the link variables in the lattice action and the
theory no longer has the nice Gaussian form that its commutative limit
does. While under certain circumstances Morita equivalence can be used
to disentangle the lattice star-product by mapping the noncommutative
lattice gauge theory onto a commutative one, the continuum limit
always requires a complicated double scaling limit to be performed
with small lattice
spacing and large commutative gauge group rank $N$, in order that the
scale of noncommutativity $\theta$ remain finite in the continuum
limit. A similar approach to solving gauge theory on the
noncommutative plane has been advocated recently in~\cite{gsv}. Nevertheless,
the lattice theory at finite $N$ can be solved
explicitly by mapping it onto a unitary two-matrix model~\cite{amns1},
whose path integral can be reduced to a well-defined sum over
integers~\cite{inprep}. This proves that the lattice model is exactly
solvable, and thereby gives a strong indication that noncommutative
gauge theory in two dimensions is a topological field theory (with no
propagating degrees of freedom).

However, such an approach, like canonical quantization in the
commutative case, is based almost entirely on the representation
theory of the gauge group. This group is a somewhat mysterious object
in noncommutative gauge theory whose full properties have not yet been
unveiled. This infinite-dimensional Lie group is analyzed
in~\cite{sheikh}--\cite{schwarzgauge} and it involves a non-trivial
mixing of colour degrees of freedom with spacetime diffeomorphisms. A
related difficulty arises in the
diagonalization approach which requires fixing a gauge symmetry
locally~\cite{blau}. The resulting Faddeev-Popov functional
determinants are difficult to analyze in the noncommutative
setting. Hamiltonian methods are likewise undesirable because of
problems associated with non-localities in time. An approach which
doesn't rely on the (unknown) features of
the noncommutative gauge group is thereby desired. We will see,
however, that the basic geometric structure underlying this gauge
group implies that the noncommutative theory is still invariant under
area-preserving diffeomorphisms of the spacetime (though in a much
stronger manner) and is thereby an exactly solvable model.

As we shall demonstrate, one technique of solving commutative $U(N)$
Yang-Mills theory which continues to be useful in the noncommutative
case is that of non-Abelian localization~\cite{Wittenloc}. This method
takes advantage of the fact that in two dimensions a gauge fixed
Yang-Mills theory is essentially a cohomological quantum field
theory. A judicious deformation of the action by cohomologically exact
terms allows one to reduce the quantum path integral defining the
partition function to a sum over a discrete set of points which are in
one-to-one correspondence with the critical points of the Yang-Mills
action. Of course, these critical points are given by gauge field
configurations which solve the classical equations of motion. Even
though these solutions may be unstable, we will refer to any such
configuration as an instanton. As a consequence, the quantum partition
function can be evaluated as a sum over all instanton configurations
of the gauge theory. In other words, the semi-classical approximation
to this field theory is {\it exact}, provided that one sums over all
critical points of the action. The feature which makes this approach
work is the interpretation of noncommutative Yang-Mills theory as
ordinary Yang-Mills theory (on a noncommutative space) with its
infinite dimensional gauge symmetry group that is formally some sort
of large $N$ limit of $U(N)$.

In what follows we will derive an exact, nonperturbative expression for the
partition function of quantum Yang-Mills theory defined on a projective
module over the noncommutative two-torus. Using a combination of
localization techniques and Morita duality, we are able to give an
explicit formula written as the sum of contributions from the
vicinity of instantons. The instantons themselves are
parameterized by a collection of lists of pairs of integers
$(\bfp,\bfq)\equiv\Bigl\{(p_k,q_k)\Bigr\}_{k\geq1}$ which arise from partitions
of
the topological numbers $(p,q)$ of the projective module
on which the gauge theory is defined. The result for the partition
function $Z_{p,q}$ is then given as a sum, over all partitions, of
terms involving the Boltzmann weights of the noncommutative Yang-Mills
action $S(\bfp,\bfq;\theta)$ evaluated at its extrema, along with
prefactors $W(\bfp,\bfq;\theta)$ which describe the quantum
fluctuations about each instanton configuration. Schematically, we have
\beq
Z_{p,q}=\sum_{\rm partitions}
W(\bfp,\bfq;\theta)~\e^{ - S(\bfp,\bfq;\theta)} \ .
\label{zschematic} \eeq
We will show that the full expression (\ref{zschematic}) is
explicitly invariant under gauge Morita equivalence and that it is a
smooth function of the noncommutativity parameter $\theta$.

The formalism which we develop in this paper gives the tools
necessary to explore and answer all questions about
two-dimensional noncommutative Yang-Mills theory, and it gives a model
which should capture some features of the more physical
higher-dimensional theories, but within a much simplified setting. For
example, the techniques developed here can be used to learn more about
the observables of Yang-Mills theory on the noncommutative torus. The
evaluation of the partition function as a sum of contributions from
instantons is of course familiar from commutative Yang-Mills
theory~\cite{dadda}--\cite{griguolo1}. In that case there exists an
equivalent expression via Poisson resummation which is interpreted as
a sum over irreducible representations of the gauge group. For
Yang-Mills theory on a noncommutative torus we have not been able to
find an analogous group theoretical expansion though we believe it
would give great insight into the representation theory of the
noncommutative gauge group on the two-dimensional torus. The
Yang-Mills action can be thought of as defining invariants of the
star-gauge group, and the discrete sums over instantons as labelling
its representations. The discrete nature of the action is necessary
for it to be a Morse function and hence a candidate for the
localization formalism~\cite{locbook}, and it suggests that the
noncommutative gauge group is compact. We expect to report on progress
in understanding the details of the noncommutative gauge group on the
torus in the near future.

\subsection{Outline and Summary of Results}

In the next section we shall begin with a review of the construction of gauge
connections and Yang-Mills theory on the two-dimensional noncommutative
torus. We include a brief discussion on the area-preserving nature of
the noncommutative gauge symmetry which suggests that there are no local
degrees of freedom in the noncommutative gauge theory, only global ones as
in the commutative case. In section~\ref{partfnloc} we give an overview
of non-Abelian localization and how
it applies to the evaluation of the quantum partition function of
two-dimensional Yang-Mills theory on the noncommutative torus. We pay
particular attention to rewriting the formalism in a manner which does
not rely on the details of the noncommutative gauge group. We show in
detail how the Yang-Mills action defines a system of Hamiltonian flows which
coincide with the Lie algebra action of the group of noncommutative gauge
transformations. This compatibility allows us to formally reduce the path
integral
defining the quantum partition function to a discrete sum.
The procedure is applicable to Yang-Mills theory defined on a noncommutative
torus with any value of the noncommutativity parameter $\theta$, including
vanishing, rational or irrational $\theta$.

The localization of the path integral
is onto gauge field configurations which are solutions of the
classical equations of motion and provide critical points of the
Yang-Mills action. In order to
characterize these solutions and the spaces in which they are defined,
in section~\ref{InstClass} we begin by giving a brief
description of finitely-generated projective (Heisenberg)
modules over the noncommutative torus. We characterize all such
classical solutions of Yang-Mills theory
defined on a projective module
for any value of the noncommutativity parameter in terms of partitions
of the topological numbers of the projective module. These results serve
to bridge previous constructions of classical solutions for
two-dimensional Yang-Mills theory in the commutative case
\cite{griguolo1,AB} and in the noncommutative case for irrational
$\theta$~\cite{rieffel}.

In order to obtain explicit results for the partition function, in
section~\ref{CYM} we revisit Yang-Mills theory on the commutative
torus and re-interpret the well-known
evaluation of the quantum partition function in this case in terms of
projective modules.  In doing so we will find it necessary to make a
distinction
between the commonly known ``physical'' definition of two-dimensional
Yang-Mills
theory and a ``module'' definition where we restrict gauge field
configurations to have a particular Chern (twist) number. The physical
theory can then be recovered by summing over all such cohomological
sectors. Given the partition function of ordinary Yang-Mills theory on
the torus written in terms of projective modules, in
section~\ref{RNCYM} we use Morita equivalence to construct a mapping from
the commutative theory to one with rational values of the noncommutativity
parameter $\theta$. Discarding the scaffolding of Morita equivalence,
the result is an explicit expression for the quantum
partition function of noncommutative Yang-Mills theory defined on a projective
module with rational $\theta$ purely by the topological numbers of
the module. Our construction also provides a more transparent interpretation
of the Morita equivalence of Yang-Mills theories on commutative tori and
ones with rational values of $\theta$.

By exploiting the fact that the localization arguments hold irrespective
of the particular value of $\theta$, in section~\ref{INCYM} we propose
a formula for the partition function at irrational values of $\theta$
by natural extension of the rational case. We give strong arguments in
favour of this conjecture. The two independent constructions of this
formula come from Morita equivalence, whereby the Morita invariant
commutative partition function determines exactly the rational
noncommutative one, and localization theory, which proves that the
partition function is given by a sum over classical solutions for any
$\theta$. Further support for this proposal is provided by rational
approximations to the irrational noncommutative gauge theory. We will
find that the schematic expression
(\ref{zschematic}) may be written explicitly as
\vspace{1cm}
\bea
Z_{p,q}&=&\sum_{\rm
  partitions}~\prod_{a\geq1}\frac{(-1)^{\nu_a}}{\nu_a!}\,
\left(\frac{g^2A}{2\pi^2}\,\Bigl(p_a-q_a\theta\Bigr)^3\right)^{-\nu_a/2}
\nn\\&&\times\,\exp\left[-\frac{2\pi^2}{g^2A}\,\sum_{k\geq1}\,
(p_k-q_k\theta)\left(\frac{q_k}{p_k-q_k\theta}-\frac
  q{p-q\theta}\right)^2\right] \ ,
\label{Zpqschemexpl}\eea
where $g$ is the Yang-Mills coupling constant and $A$ is the area of
the torus. The integer $\nu_a$ is the number of partition components
$(p_k,q_k)$ which have the same distinct values of the quantity
$p_a-q_a\theta$. The sign factor in (\ref{Zpqschemexpl}) is determined
by a Morse index which measures the overall contribution from unstable
modes in a given instanton configuration $(\bfp,\bfq)$. The exponential
prefactors are the Gaussian fluctuation determinants, weighted with
the appropriate permutation symmetry factors $\nu_a!$ associated with
a partition. From (\ref{Zpqschemexpl}) we see that the area dependence
of the noncommutative gauge theory is similar to that of the
commutative case. If $A\to\infty$ for fixed $g$ and $\theta$, then the
theory is exponentially dominated by trivial instanton
configurations. Essentially the energy of electric flux in the
noncommutative theory is still proportional to the length of the flux
line, and so the overall details of the dynamics (or lack thereof) are
the same as in commutative Yang-Mills theory. Thus, in direct analogy
to the commutative situation, the gauge theory on the noncommutative
plane is essentially trivial.

In section~\ref{smooth} we develop a graphical method of analyzing the
instanton contributions to Yang-Mills theory which works for
$\theta$ irrational, rational or vanishing. This graphical approach is
applied to the universal expression (\ref{Zpqschemexpl}) for the
partition function to show that the vacuum energy, along with a
certain class of topological observables, of Yang-Mills theory on the
noncommutative torus are smooth functions of $\theta$. Finally, in
section~\ref{moduli} we end with a description of the moduli spaces of
classical solutions of Yang-Mills theory on the noncommutative
torus. The partition function in the weak coupling limit agrees with
that of the commutative gauge theory, except that now it formally
computes the symplectic volume of the moduli space of {\it all} (not
necessarily flat) constant curvature gauge connections on the
torus. The rearrangement of the series (\ref{Zpqschemexpl}) into
distinct gauge inequivalent instanton configurations is
described. They are determined by rearranging the critical partition
components $(p_k,q_k)$ into distinct relatively prime pairs
$(p_a',q_a')$ of topological numbers with
$(p_a,q_a)=N_a\,(p_a',q_a')$. We will see that the moduli space of such
gauge orbits is given by
\beq
{\cal M}_{p,q}=\prod_{a\geq1}\,{\rm Sym}^{N_a}\,\tilde{\bf T}^2 \ ,
\label{calMpqschem}\eeq
where ${\rm Sym}^{N_a}\,\tilde{\bf T}^2$ is the symmetric product of a
certain dual, ordinary two-torus $\tilde{\bf T}^2$. This generalizes the moduli
space ${\rm Sym}^N\,\tilde{\bf T}^2$ of flat gauge connections in
commutative $U(N)$ gauge theory. The instanton moduli space
(\ref{calMpqschem}) has a natural physical interpretation in terms of
that for a collection of distinct configurations of $N_a$ free
indistinguishable
D0-branes in codimension two. In particular, the point-like
instanton singularities are not resolved by noncommutativity. We will show
how the orbifold singularities of (\ref{calMpqschem}) can be used to
systematically construct the gauge inequivalent contributions to
Yang-Mills theory. Such an explicit classification is only possible
within the noncommutative setting. We shall find that, like for the
instanton contributions to ordinary Yang-Mills theory, there are a
finite number of quantum fluctuations about each gauge inequivalent
classical solution. In contrast to the commutative case, however, for
irrational $\theta$ there are infinitely many distinct instanton
contributions to the path integral for fixed quantum numbers $(p,q)$.

\newsection{Noncommutative Gauge Theory in Two Dimensions\label{2dncym}}

To set notation and conventions, we will start by reviewing some well-known
facts about Yang-Mills theory on a noncommutative
two-torus~\cite{ks,cr,connesbook}. Our presentation
will exhibit the interplay between the physical, quantum field theoretical
approach and the mathematical approach within the framework of noncommutative
geometry, as both descriptions will be fruitful for our subsequent analysis in
later sections. We will also give the first indication that this theory is
exactly solvable. For simplicity, we consider a square torus of radii $R$.

\subsection{The Noncommutative Torus\label{nctorus}}

The noncommutative two-torus may be defined as the abstract, noncommutative,
associative unital $*$-algebra generated by two unitary operators $\hat Z_1$
and $\hat Z_2$ with the commutation relation
\beq
\hat Z_1\hat Z_2=\e^{2\pi\ii\theta}~\hat Z_2\hat Z_1 \ ,
\label{NCcommrel}\eeq
where $\theta$ is the real-valued, dimensionless noncommutativity parameter.
Unless otherwise specified, we will assume that $\theta\in(0,1)$ is an
irrational number. The ``smooth'' completion $\alg_\theta$ of the algebra
generated by $\hat Z_1$ and $\hat Z_2$ consists of the power series
\beq
\hat f=\sum_{m_1=-\infty}^\infty~
\sum_{m_2=-\infty}^\infty f^{~}_{(m_1,m_2)}~\e^{\pi\ii\theta\,
m_1m_2}~\hat Z_1^{m_1}\hat Z_2^{m_2} \ ,
\label{hatf}\eeq
where the coefficients $f^{~}_{(m_1,m_2)}$ are Schwartz functions of
$(m_1,m_2)\in\zed^2$, i.e. $f^{~}_{(m_1,m_2)}\to0$ faster than any power of
$|m_1|+|m_2|$ as $|m_1|+|m_2|\to\infty$. The phase factor in (\ref{hatf}) is
inserted to symmetrically order the operator product.

There are natural, anti-Hermitian linear derivations $\hat\partial_1$ and
$\hat\partial_2$ of the algebra $\alg_\theta$ which are defined by the
commutation relations
\bea
\left[\hat\partial_1\,,\,\hat\partial_2\right]&=&\ii\Phi\cdot\id \ ,
\label{Heisen}\\\left[\hat\partial_i\,,\,\hat Z_j\right]&=&
\frac{\ii}{R}\,\delta_{ij}\,\hat Z_j \ , ~~ i,j=1,2 \ ,
\label{partialderiv}\eeq
where $\Phi\in\real$ can be interpreted as a background magnetic flux
and $\id$ is the unit of $\alg_\theta$. From
(\ref{Heisen}) it follows that the Heisenberg Lie algebra ${\cal L}_\Phi$ acts
on $\alg_\theta$ by infinitesimal automorphisms. This action defines a Lie
algebra homomorphism $X\mapsto\hat\partial_X$, $X\in{\cal L}_\Phi$, i.e.
$\left[\hat\partial_X\,,\,\hat\partial_Y\right]=\hat\partial_{[X,Y]}$,
yielding a linear map
\beq
\hat\partial\,:\,\alg_\theta~\longrightarrow~\alg_\theta\otimes{\cal
  L}_\Phi^* \ .
\label{partialmap}\eeq
The unique normalized trace on $\alg_\theta$ is given by projection
onto zero modes as
\beq
\Tr\,\hat f=f^{~}_{(0,0)} \ ,
\label{Trdef}\eeq
which defines a positive linear functional $\alg_\theta\to\complex$, i.e.
$\Tr\,\hat f^\dag\hat f\geq0$ for any $\hat f\in\alg_\theta$. The trace
(\ref{Trdef}) satisfies $\Tr\,\hat f^\dag=\overline{\Tr\,\hat f}$, and it is
invariant under the action of the Lie algebra ${\cal L}_\Phi$ of automorphisms
of $\alg_\theta$, i.e.
\beq
\Tr\left[\hat\partial_i\,,\,\hat f\,\right]=0 \ .
\label{Trautoinv}\eeq

The conventional field theoretic approach employs a ``dual'' description to
this analytic one in terms of functions on an ordinary torus ${\bf T}^2$. Let
$x^1,x^2\in[0,2\pi R]$ be the coordinates of ${\bf T}^2$. Then given any
element $\hat f\in\alg_\theta$ with series expansion of the form (\ref{hatf}),
we can use the Schwartz sequence $f^{~}_{(m_1,m_2)}$ to define a smooth
function on the torus by the Fourier series
\beq
f(x)=\sum_{m_1=-\infty}^\infty~\sum_{m_2=-\infty}^\infty f^{~}_{(m_1,m_2)}~
\e^{\ii m_ix^i/R} \ .
\label{fFourier}\eeq
This establishes a one-to-one correspondence between elements of the abstract
algebra $\alg_\theta$ and elements of the algebra $C^\infty({\bf T}^2)$ of
smooth functions on the torus. Under this correspondence, the noncommutativity
of $\alg_\theta$ is encoded in the multiplication relation
\beq
\hat f\,\hat g=\widehat{f\star g} \ ,
\label{fgprod}\eeq
where the star-product is given by
\beq
(f\star g)(x)=\sum_{n=0}^\infty\left(-\pi\ii R^2\theta\right)^n\,
\sum_{r=0}^n\frac{(-1)^r}{(n-r)!\,r!}\,\Bigl(\partial_1^r\,\partial_2^{n-r}
f(x)\Bigr)\Bigl(\partial_1^{n-r}\,\partial_2^rg(x)\Bigr)
\label{stardef}\eeq
with $\partial_i=\partial/\partial x^i$. In addition, the actions of the
derivations (\ref{partialderiv}) correspond to ordinary differentiation
of functions,
\beq
\left[\hat\partial_i\,,\,\hat f\,\right]=\widehat{\partial_if} \ ,
\label{partialf}\eeq
while the canonical normalized trace (\ref{Trdef}) can be represented in terms
of the classical average of functions over the torus,
\beq
\Tr\,\hat f=\frac1{4\pi^2R^2}\,\int\dd^2x~f(x) \ .
\label{Travg}\eeq
Integration by parts also shows that
\beq
\int\dd^2x~(f\star g)(x)=\int\dd^2x~f(x)\,g(x) \ .
\label{star2}\eeq
Here and in the following, unless specified otherwise, all coordinate
integrations extend over ${\bf T}^2$.

\subsection{Gauge Theory on the Noncommutative Torus\label{nctorusym}}

In the noncommutative setting, the generalizations of vector bundles are
provided by projective modules, which are vector spaces on which the algebra is
represented. Let $\cal E$ be a finitely-generated projective module over the
algebra $\alg_\theta$. We consider only right modules in the following. The
free module $\alg_\theta^M=\alg_\theta\oplus\cdots\oplus\alg_\theta$ consists
of $M$-tuples $\hat\xi=(\hat f_1,\dots,\hat f_M)$ of elements $\hat
f_a\in\alg_\theta$. It is the analog of a trivial vector bundle. Let
${\sf P}\in\mat_M(\alg_\theta)$ be a projector with
\beq
{\cal E}={\sf P}\,\alg_\theta^M \ , ~~ {\sf P}^2={\sf P}={\sf P}^\dag
\ ,
\label{calEPAM}\eeq
where $\mat_M(\alg_\theta)=\alg_\theta\otimes\mat_M$ is the algebra of $M\times
M$
matrices with entries in the algebra $\alg_\theta$, whose multiplication is the
tensor product of the multiplication in $\alg_\theta$ with ordinary matrix
multiplication. Alternatively, we may consider $\cal E$ as the subspace of
elements $\hat\xi\in\alg_\theta^M$ with ${\sf P}\,\hat\xi=\hat\xi$.

The endomorphism algebra ${\rm End}_{\alg_\theta}({\cal E})={\cal E}^*
\otimes_{{\cal A}_\theta}{\cal E}$ of the module
$\cal E$ is the algebra of linear maps ${\cal E}\to{\cal E}$ that commute with
the right action of $\alg_\theta$ on $\cal E$. It is isomorphic to the
subalgebra of $\alg_\theta$-valued matrices $\hat A\in\mat_M(\alg_\theta)$
which obey ${\sf P}\,\hat A\,{\sf P}=\hat A$. This means that the identity
operator on ${\cal E}$ can be identified with the projector, $\id_{\cal
E}={\sf P}$. To simplify some of the formulas which follow, we shall frequently
refrain from writing $\id_{\cal E}$ explicitly. Let $N\leq M$ be the largest
integer such that the module $\cal E$ can be represented as a direct sum ${\cal
E}={\cal E}'\oplus\cdots\oplus{\cal E}'$ of $N$ isomorphic
$\alg_\theta$-modules. Then ${\rm End}_{\alg_\theta}({\cal
E}')\cong\alg_{\theta'}$ is also a noncommutative torus~\cite{cr},
where $\theta'$ is the {\it dual} noncommutativity parameter which
depends on $\theta$ and the projective module $\cal E$, so that
\beq
{\rm End}_{\alg_\theta}({\cal E})\cong\mat_N(\alg_{\theta'}) \ .
\label{EndMN}\eeq

The derivations $\hat\partial_i$ naturally extend to operators on
$\alg_\theta^M$ via the definition
\beq
\left[\hat\partial_i\,,\,\hat\xi\,\right]=\left(\left[\hat\partial_i\,,\,
\hat f_1\right]\,,\,\dots\,,\,\left[\hat\partial_i\,,\,\hat f_M\right]\right)
\label{partialicalE}\eeq
for $\hat\xi=(\hat f_1,\dots,\hat f_M)\in\alg_\theta^M$. Then ${\sf
  P}\circ\hat\partial_i\circ{\sf P}$ is a linear derivation on $\cal
  E$. The trace Tr on $\alg_\theta$ also naturally extends to a trace on ${\rm
End}_{\alg_\theta}({\cal E})$ defined by
\beq
\Tr^{~}_{\cal E}=\Tr\otimes\tr^{~}_M \ ,
\label{TrcalEdef}\eeq
where $\tr^{~}_M$ is the usual $M\times M$ matrix trace. On $\cal E$ there
is a natural $\alg_\theta$-valued inner product which is compatible with the
$\alg_\theta$-module structure of $\cal E$ and is defined on $M$-tuples
$\hat\xi=(\hat f_1,\dots,\hat f_M)$ and $\hat\eta=(\hat g_1,\dots,\hat g_M)$
by
\beq
\left\langle\hat\xi\,,\,\hat\eta\right\rangle^{~}_{\alg_\theta}=
\sum_{a=1}^M\hat f_a^\dag\,\hat g_a \ .
\label{calEinnerprod}\eeq
The object
\beq
\left\langle\hat\xi\,,\,\hat\eta\right\rangle=\Tr\left
\langle\hat\xi\,,\,\hat\eta\right\rangle^{~}_{\alg_\theta}
\label{ordinnerprod}\eeq
then defines an ordinary Hermitian scalar product ${\cal E}\times{\cal
E}\to\complex$. This turns $\cal E$ into a separable Hilbert space. We will
present the explicit classification of the projective modules over the
noncommutative torus in section~\ref{Heisenberg}.

We now define a connection on a module $\cal E$ over the noncommutative torus
to be a pair of linear operators $\hat\nabla_1,\hat\nabla_2:{\cal E}\to{\cal
E}$ satisfying
\beq
\left[\hat\nabla_i\,,\,\hat Z_j\right]=
\frac{\ii}{R}\,\delta_{ij}\,\hat Z_j \ , ~~ i,j=1,2 \ ,
\label{nablaidef}\eeq
where in this equation the $\hat Z_j$ are regarded as operators ${\cal
E}\to{\cal E}$ representing the right action on $\cal E$ of the corresponding
generators of $\alg_\theta$. When acting on elements of $\cal E$, the
requirement (\ref{nablaidef}) is just the usual Leibnitz rule with respect to
the derivations $\hat\partial_1$ and $\hat\partial_2$. In an analogous way to
these operators, there is a linear map $X\mapsto\hat\nabla_X$, $X\in{\cal
L}_\Phi$, which defines a vector space homomorphism
\beq
\hat\nabla\,:\,{\cal E}~\longrightarrow~{\cal
  E}\otimes_\complexs{\cal L}_\Phi^* \ .
\label{nablamap}\eeq
This definition makes use of the bimodule structure on
$\alg_\theta\otimes{\cal L}_\Phi^*$. From the definitions
(\ref{partialderiv}) and (\ref{nablaidef}) it
follows that an arbitrary connection $\hat\nabla_i$ can be expressed in the
form
\beq
\hat\nabla_i=\hat\partial_i+\hat A_i \ ,
\label{gaugefielddef}\eeq
where $\hat A_i\in{\rm End}_{\alg_\theta}({\cal E})$ are $N\times N$
$\alg_{\theta'}$-valued matrices which we will refer to as gauge fields. We
stress that here, and below, the quantity $\hat\partial_i$ is implicitly
understood as the operator ${\sf P}\circ\hat\partial_i\circ{\sf P}$ on
$\alg_\theta^M\to\cal E$. The same is true of similarly defined objects.

In the following we shall work only with connections which are compatible with
the inner product (\ref{calEinnerprod}), i.e. those which satisfy
\beq
\left\langle\hat\nabla_i\,\hat\xi\,,\,\hat\eta\right
\rangle_{\alg_\theta}+\left\langle\hat
\xi\,,\,\hat\nabla_i\,\hat\eta\right\rangle_{\alg_\theta}=
\left[\hat\partial_i~,~\left\langle
\hat\xi\,,\,\hat\eta\right\rangle_{\alg_\theta}\right]
\label{compconn}\eeq
for any $\hat\xi,\hat\eta\in{\cal E}$. The compatibility condition
(\ref{compconn}) implies that $\hat\nabla_i$ is an anti-Hermitian operator with
respect to the scalar product (\ref{ordinnerprod}). It also implies that its
curvature $\left[\hat\nabla_1\,,\,\hat\nabla_2\right]$, which is a two-form on
the Heisenberg algebra ${\cal L}_\Phi$ with values in the space of linear
operators on $\cal E$, commutes with the action of $\alg_\theta$ on $\cal E$,
and hence takes values in the space ${\rm End}_{\alg_\theta}^{\rm H}({\cal E})$
of anti-Hermitian endomorphisms of ${\cal E}$.\footnote{\baselineskip=12pt
Usually one would define the curvature to be a measure of the deviation of the
mapping $X\mapsto\hat\nabla_X$ from being a homomorphism of the Lie algebra
(\ref{Heisen}) of automorphisms of $\alg_\theta$. This means that the curvature
should be defined as
$\left[\hat\nabla_1\,,\,\hat\nabla_2\right]-\Phi\cdot\id_{\cal E}$. However,
later on we will wish to work with an action which is explicitly invariant
under Morita duality, which can only be accomplished with the definition of
curvature given in the text. This change of convention is mathematically
harmless since it corresponds to a shift of the curvature by the central
element of the Heisenberg algebra. Physically, it will only add constants to
the usual gauge theory action and so will not affect any local dynamics, only
topological aspects.} The space of all compatible connections on a module $\cal
E$ will be denoted by $\CC({\cal E})$. From (\ref{nablaidef}) and
(\ref{compconn}) it follows that $\CC({\cal E})$ is an affine space over the
vector space of linear maps ${\cal L}_\Phi\to{\rm End}_{\alg_\theta}^{\rm
H}({\cal E})$.

In this paper we will be interested in evaluating the partition function of two
dimensional quantum Yang-Mills theory on the noncommutative torus, which is
defined formally by the infinite-dimensional integral
\beq
Z(g^2,\theta,\Phi,{\cal E})=
\frac1{{\rm vol}\,{\sf G}({\cal E})}~\int\limits_{\CC({\cal E})}
\DD\hat A~\e^{-S\bigl[\hat A\bigr]} \ ,
\label{ZNCYMop}\eeq
where the Yang-Mills action on $\CC({\cal E})$ is defined for an arbitrary
connection (\ref{gaugefielddef}) by
\beq
S\left[\hat A\right]=S\left[\hat\nabla\right]=\frac{2\pi^2R^2}
{g^2}\,\Tr^{~}_{\cal E}\left[\hat\nabla_1\,,\,\hat\nabla_2\right]^2
\label{SNCYMop}\eeq
with $g$ the Yang-Mills coupling constant of unit mass dimension. The
area factor $4\pi^2R^2$ is inserted to make the action dimensionless. Here
${\sf G}({\cal E})$ is the group of gauge transformations, which will be
described in the next subsection, and ${\rm vol}\,{\sf G}({\cal E})$ is
its volume. The measure $\DD\hat A$, and also the volume ${\rm vol}\,{\sf
G}({\cal E})$, will be defined more precisely in section~\ref{partfnloc}. By
using the operator-field correspondence of the previous
subsection, we can express (\ref{ZNCYMop}) in a more standard quantum field
theoretical form as the Euclidean Feynman path integral
\beq
Z(g^2,\theta,\Phi,{\cal E})=
\frac1{{\rm vol}\,{\sf G}({\cal E})}~\int\limits_{\CC({\cal E})}
\DD A~\e^{-S[A]} \ ,
\label{ZNCYMfield}\eeq
where
\beq
S[A]=\frac1{2g^2}\,
\int\dd^2x~\tr^{~}_N\Bigl(F_A(x)+\Phi\cdot\id_{\cal E}\Bigr)^2
\label{SNCYMfield}\eeq
with
\beq
F_A=\partial_1A_2-\partial_2A_1+A_1\star'A_2-A_2\star'A_1
\label{FAdef}\eeq
the noncommutative field strength of the anti-Hermitian $U(N)$ gauge field
$A_i$. The multiplication in (\ref{FAdef}) is the tensor product of the
associative star-product (\ref{stardef}), defined with $\theta$ replaced by its
dual $\theta'$, and ordinary matrix multiplication. This extended star-product
is still associative.

\subsection{Gauge Symmetry and Area Preserving Diffeomorphisms\label{gaugesym}}

Let us now describe the symmetries of the noncommutative Yang-Mills action
(\ref{SNCYMop}). It is invariant under any covariant transformation of the
gauge connection of the form
\beq
\hat\nabla_i~\longmapsto~\hat U\,\hat\nabla_i\,\hat U^\dag \ ,
\label{nablatransf}\eeq
where $\hat U\in{\rm End}_{\alg_\theta}({\cal E})$ is a unitary endomorphism of
the projective module $\cal E$,
\beq
\hat U^\dag\,\hat U=\hat U\,\hat U^\dag=\id_{\cal E} \ ,
\label{hatUunitary}\eeq
which determines an inner automorphism of the right action of $\alg_\theta$ on
$\cal E$. In other words, $\hat U\in\unitary_N(\alg_{\theta'})$, where
$\unitary_N(\alg_{\theta'})$ is the group of unitary elements of the
algebra $\mat_N(\alg_{\theta'})$. These gauge transformations comprise
operators of the form
\beq
\hat U=\id_{\cal E}+\hat K \ ,
\label{unitaryend}\eeq
where $\hat K$ lies in an appropriate completion of the algebra of finite rank
endomorphisms of $\cal E$~\cite{nair,harvey}. These latter
endomorphisms are defined as follows.
For any $\hat\eta,\hat\eta'\in{\cal E}$, let
$\left|\hat\eta\right\rangle\left\langle\hat\eta'\right|$ be the operator
defined by
\beq
\left|\hat\eta\right\rangle\left\langle\hat\eta'\right|\hat\xi=
\hat\eta\,\left\langle\hat\eta'\,,\,\hat\xi\,\right\rangle^{~}_{\alg_\theta}
\label{finiterankdef}\eeq
for $\hat\xi\in{\cal E}$, with adjoint
$\left|\hat\eta'\,\right\rangle\left\langle\hat\eta\right|$. The
$\alg_\theta$-linear span of endomorphisms of the form
(\ref{finiterankdef}) forms a self-adjoint two-sided ideal in ${\rm
  End}_{\alg_\theta}({\cal E})$. Since, as mentioned before, $\cal E$
is a separable Hilbert space, this ideal is isomorphic to the
infinite-dimensional algebra $\mat_\infty$ of finite rank matrices. Its
operator norm closure is the algebra ${\rm End}_{\alg_\theta}^\infty({\cal E})$
of compact endomorphisms of the module $\cal E$.

The Schwartz restriction on the expansion (\ref{hatf}) implies that elements
$\hat f\in\alg_\theta$ act as compact operators on $\cal E$~\cite{LSZ}.
Therefore, in
(\ref{nablatransf}) we should restrict to those unitary endomorphisms
(\ref{unitaryend}) with $\hat K\in{\rm End}_{\alg_\theta}^\infty({\cal E})$. We
denote this infinite dimensional Lie group by $U^\infty({\cal E})$. It is the
operator norm completion of the infinite unitary group $U(\infty)$ obtained by
taking $\hat K$ to be a finite rank endomorphism.\footnote{\baselineskip=12pt
The gauge group can also be chosen to be smaller than $U^\infty({\cal E})$ by
completing $U(\infty)$ in other Schatten
norms~\cite{harvey}--\cite{schwarzgauge}. The various choices
all have the same topology and group theory, and so we shall work for
definiteness with only the compact unitaries defined above.} By
Palais' theorem~\cite{palais}, these two
unitary groups have the same homotopy type, and their homotopy groups are
determined by Bott periodicity as
\beq
\pi_k^{~}\Bigl(U^\infty({\cal E})\Bigr)=\pi_k^{~}
\Bigl(U(\infty)\Bigr)=\left\{\new
{\begin{array}{lll}\zed \ &,& ~~ k~{\rm odd} \ , \\0 \ &,& ~~
k~{\rm even} \ . \end{array}}\right.
\label{gaugehomotopy}\eeq
In particular, the gauge symmetry group is connected. It should be pointed out
here that this is only a {\it local} description of the full gauge group of
noncommutative Yang-Mills theory. The group of connected components of
${\sf G}({\cal E})$ acts on the gauge orbit space, obtained by quotienting
$\CC({\cal E})$ by the action of the group ${\sf G}_0({\cal E})$ of smooth
maps ${\bf T}^2\to U^\infty({\cal E})$, as a global symmetry
group~\cite{ks,harvey,schwarzgauge}.

By using (\ref{gaugefielddef}), one finds that the infinitesimal form of the
gauge transformation rule (\ref{nablatransf}) is $\hat A_i\mapsto\hat
A_i+\delta_{\hat\lambda}\hat A_i$, where
\beq
\delta_{\hat\lambda}\hat A_i=-\left[\hat\partial_i\,,\,\hat\lambda\right]+
\left[\hat\lambda\,,\,\hat A_i\right]
\label{gaugetransfop}\eeq
and $\hat\lambda$ is an anti-Hermitian compact operator on $\cal E$. In terms
of gauge potentials on the ordinary torus ${\bf T}^2$ this reads
\beq
\delta_\lambda A_i=-\partial_i\lambda+\lambda\star'A_i-A_i\star'\lambda \ ,
\label{gaugetransffield}\eeq
where $\lambda(x)$ is a smooth, anti-Hermitian $N\times N$ matrix-valued field
on ${\bf T}^2$. The noncommutative gauge transformations
(\ref{gaugetransffield}) mix internal, $U(N)$ gauge degrees of freedom with
general coordinate transformations of the torus. Their geometrical significance
has been elucidated in~\cite{LSZ} by exploiting the relationship between
appropriate completions of $U(\infty)$ and canonical transformations. The Lie
algebra of noncommutative gauge transformations (\ref{gaugetransffield}) is
equivalent to the Fairlie-Fletcher-Zachos trigonometric
deformation~\cite{FFZ} of the
algebra $w_\infty({\bf T}^2)$ of area-preserving diffeomorphisms of ${\bf
T}^2$.

Therefore, the gauge symmetry group of noncommutative Yang-Mills theory in
two-dimensions consists of area-preserving diffeomorphisms, which ``almost''
makes it a topological field theory. Its gauge symmetry ``almost'' coincides
with general covariance, thereby killing most of its degrees of freedom. From
this feature we would expect the theory to contain no local propagating degrees
of freedom, and hence to be exactly solvable. This reasoning is further
supported by the Seiberg-Witten map~\cite{sw} and the exact solvability of
ordinary,
commutative Yang-Mills gauge theory in two dimensions. Note however that the
topological nature here is quite different than that of the commutative case,
because in the noncommutative setting it arises due to the {\it gauge symmetry}
of the theory, i.e. an inner automorphism of the algebra of functions, while in
the commutative case it corresponds to an outer automorphism which preserves
the local area element $4\pi^2R^2~\dd^2x$. For this reason, the partition
function will only depend on the dimensionless combination $4\pi^2\,g^2\,R^2$
of the Yang-Mills coupling constant and the area of the surface. This fact
makes it difficult to make sense of the theory on a non-compact surface.

In contrast, this argument breaks down for noncommutative tori of dimension
larger than two. In any even dimension the transformations
(\ref{gaugetransffield}) generate symplectic
diffeomorphisms~\cite{LSZ}, i.e. coordinate transformations which
leave the symplectic two-form of the torus invariant. These are the
diffeomorphisms which preserve the Poisson bi-vector defining the star-product
in (\ref{stardef}). In general, these transformations generate a group that is
much smaller than the group of volume-preserving diffeomorphisms. In the
D-brane interpretation, this latter group would be the natural worldvolume
symmetry group of a static brane. However, particular to the two-dimensional
case is the fact that canonical transformations and area-preserving
diffeomorphisms are the same.

\newsection{Localization of the Partition Function\label{partfnloc}}

The  path integral (\ref{ZNCYMop}) for quantum Yang-Mills theory on the
noncommutative torus has several features in common with
non-Abelian gauge theory defined on an ordinary, commutative
torus. Formally, it can be regarded as a certain ``large
$N$ limit'' of ordinary $U(N)$ Yang-Mills where we have generalized the gauge
fields to measurable operators. In this
section we will exploit these similarities to show how one may compute exactly
the partition function for noncommutative gauge theory on the torus
via the technique of non-Abelian localization~\cite{Wittenloc}. The first key
observation we shall make is that the integration measure $\DD\hat A$ in
(\ref{ZNCYMop}) may be naturally identified with the gauge invariant Liouville
measure induced on the infinite dimensional operator space of compatible
connections $\CC({\cal E})$ by a symplectic two-form $\omega[\cdot,\cdot]$.
Moreover, the volume of the gauge group ${\rm vol}\,{\sf G}({\cal E})$ is
determined formally from the volume form on ${\sf G}({\cal E})$ associated
with the metric $\left(\hat\lambda\,,\,\hat\lambda\right)=\Tr^{~}_{\cal
E}\,\hat\lambda^2$ on ${\rm End}_{\alg_\theta}^{{\rm
H},\infty}({\cal E})$. This metric also induces an invariant
quadratic form $(\cdot,\cdot)$ on the dual Lie algebra $\Bigl({\rm
End}_{\alg_\theta}^{{\rm H},\infty}({\cal E})\Bigr)^*$, such that
the noncommutative Yang-Mills action (\ref{SNCYMop}) is
proportional to the square of the moment map $\mu$ corresponding to the
symplectic action of ${\sf G}({\cal E})$ on $\CC({\cal E})$.
Equivalently, the Lie algebra action of the group of gauge
transformations ${\sf G}({\cal E})$ coincides with a
system of Hamiltonian flows defined by the Yang-Mills action.
In particular, this implies that the action (\ref{SNCYMop}) is a
gauge-equivariant Morse function on~$\CC({\cal E})$~\cite{AB,rieffel}.

Consequently, the partition function of Yang-Mills theory defined on the
noncommutative torus can be expressed formally as an infinite-dimensional
statistical mechanics model
\beq
Z(g^2,\theta,\Phi,{\cal E})= \frac1{
{\rm vol}\,{\sf G}({\cal E})}~\int\limits_{\CC({\cal E})}\exp
\left[\omega-\frac{\beta}{2}\,\Bigl(\mu\,,\,\mu\Bigr)\right] \ ,
\label{statmechZ}
\eeq
where
\beq
\beta=\frac{4\pi^2R^2}{g^2} \ .
\label{betadef}\eeq
As shown in~\cite{Wittenloc}, path integrals of the form (\ref{statmechZ}) are
formally calculable through a generalized non-Abelian localization technique.
Here ``localization'' refers to the fact that the path integral
(\ref{statmechZ}) is given exactly by the sum over contributions from
neighbourhoods of stationary points of the Yang-Mills action
(\ref{SNCYMop}). If we
denote the discrete set of all such critical points by ${\cal P}(\theta,{\cal
E})$, then
\beq
Z(g^2,\theta,\Phi,{\cal E})= \sum_{\hat A^{\rm cl}\in{\cal P}(\theta,{\cal E})}
W\left[\hat A^{\rm cl}\right]~\e^{-\frac{\beta}{2}\,\bigl(\mu
[\hat A^{\rm cl}]\,,\,\mu[\hat A^{\rm cl}]\bigr)} \ ,
\label{ZNCYMcritsum}\eeq
where the function $W$ gives the contributions due to the quantum fluctuations
about
the stationary points. In the remainder of this section we will derive all of
these properties in some detail.

\subsection{Symplectic Structure\label{sympl}}

Let $\cal E$ be a finitely-generated projective module over the noncommutative
torus, and consider the space $\CC({\cal E})$ of compatible connections on
$\cal E$ introduced in section~\ref{nctorusym}. The group of gauge
transformations ${\sf G}({\cal E})$ acts on $\CC({\cal E})$ and it has Lie
algebra ${\rm End}_{\alg_\theta}^{{\rm H},\infty}({\cal E})$ consisting of
anti-Hermitian compact operators on $\cal E$. On this Lie algebra we introduce
a natural invariant, non-degenerate quadratic form by
\beq
\left(\hat\lambda\,,\,\hat\lambda'\right)=\Tr^{~}_{\cal E}\,\hat\lambda\,
\hat\lambda' \ , ~~ \hat\lambda,\hat\lambda'\in
{\rm End}_{\alg_\theta}^{{\rm H},\infty}({\cal E}) \ .
\label{Endquad}\eeq
The infinitesimal gauge transformations (\ref{gaugetransfop}) define a
group action on $\CC({\cal E})$ because
\beq
\left[\delta_{\hat\lambda}\,,\,\delta_{\hat\lambda'}\right]\hat A_i=
\delta_{\bigl[\hat\lambda\,,\,\hat\lambda'\bigr]}\hat A_i \ .
\label{gpaction}\eeq
Consider the representation (\ref{Heisen}) of the Heisenberg algebra ${\cal
L}_\Phi$ in the Lie algebra of derivations of $\alg_\theta$, and let
$\Lambda({\cal L}_\Phi^*)=\bigoplus_{n\geq0}\Lambda^n({\cal L}_\Phi^*)$ be the
$\zed_+$-graded exterior algebra of ${\cal L}_\Phi$. To this representation
there corresponds the graded differential algebra
\beq
\Omega({\cal E})=\bigoplus_{n\geq0}\Omega^n({\cal E}) \ , ~~
\Omega^n({\cal E})={\rm End}_{\alg_\theta}^{{\rm H},\infty}({\cal E})
\otimes_\complexs\Lambda^n\left({\cal L}_\Phi^*\right)
\label{OmegacalEdef}\eeq
of left-invariant differential forms on $\exp({\cal L}_\Phi)$ with coefficients
in ${\rm End}_{\alg_\theta}^{{\rm H},\infty}({\cal E})$. For instance, the
curvature $\left[\hat\nabla_1\,,\,\hat\nabla_2\right]=\hat F_{\hat
A}+\Phi\cdot\id_{\cal E}\in\Omega^2({\cal E})$ with ${\sf G}({\cal E})$
acting infinitesimally through
\beq
\delta_{\hat \lambda}\hat F_{\hat
A}=\left[\hat\lambda\,,\,\hat F_{\hat A}\right]
\label{deltalambdaF}\eeq
with the usual product on the differential algebra
  (\ref{OmegacalEdef}) implicitly understood. Functional
  differentiation at a point $\hat A\in\CC({\cal E})$ is then defined
  through
\beq
\frac{\delta}{\delta\hat A}f\left[\hat a\right]\equiv\left.\frac
\dd{\dd t}f\left[\hat A+t\,\hat a\right]\right|_{t=0} \ , ~~ \hat
a\in\Omega^1({\cal E}) \ .
\label{fnderivdef}\eeq

As mentioned in section~\ref{nctorusym}, $\CC({\cal E})$ is an affine space
over the vector space ${\rm End}_{\alg_\theta}^{{\rm H},\infty}({\cal
E})\otimes_\complexs{\cal L}_\Phi^*$ of linear maps ${\cal L}_\Phi\to{\rm
End}_{\alg_\theta}^{{\rm H},\infty}({\cal E})$, whose tangent space can be
identified with the cotangent space ${\rm End}_{\alg_\theta}^{{\rm
H},\infty}({\cal E})\otimes_\complexs\Lambda^1({\cal
L}_\Phi^*)=\Omega^1({\cal E})$. A
natural symplectic structure may then be defined on $\CC({\cal E})$ by the
two-form
\beq
\omega\left[\hat a,\hat a'\right]=\Tr_{\cal E}^{~}\,\hat a\wedge
\hat a' \ , ~~ \hat a,\hat a'\in\Omega^1({\cal E}) \ ,
\label{omegadef}\eeq
where\footnote{\baselineskip=12pt Note that components associated with
  the central elements of the Heisenberg algebra ${\cal L}_\Phi$
  trivially drop out of all formulas such as (\ref{wedgedef}), and hence
  will not be explicitly written in what follows.}
\beq
\hat a\wedge\hat a'=\hat a_1\,\hat a_2'-\hat a_2\,\hat a_1' \ .
\label{wedgedef}\eeq
Since (\ref{omegadef}) is independent of the point $\hat A\in\CC({\cal
  E})$ at which it is evaluated, it is closed,
i.e. $\delta\omega/\delta\hat A=0$, and it is also clearly
non-degenerate. In fact, because of the identities (\ref{Travg}),
(\ref{star2}) and (\ref{TrcalEdef}), the symplectic two-form
(\ref{omegadef}) coincides with the canonical, commutative one that is
usually introduced in ordinary two-dimensional $U(N)$ Yang-Mills
  theory~\cite{AB}. Its main characteristic is that it
is invariant under the infinitesimal action
\beq
\delta_{\hat\lambda}\hat a_i=\left[\hat\lambda\,,\,\hat a_i\right]
\label{deltalambdaai}\eeq
of the gauge group ${\sf G}({\cal E})$ on $\CC({\cal E})$,
\beq
\omega\left[\hat a,\delta_{\hat\lambda}\hat a'\right]+
\omega\left[\delta_{\hat\lambda}\hat a,\hat a'\right]=0 \ .
\label{omegainv}\eeq

\subsection{Hamiltonian Structure\label{momap}}

Since $\CC({\cal E})$ is contractible and ${\sf G}({\cal E})$ acts
symplectically on $\CC({\cal E})$ with respect to the symplectic structure
(\ref{omegadef}), there exists a moment map
\beq
\mu\,:\,\CC({\cal E})~\longrightarrow~\Bigl(
{\rm End}_{\alg_\theta}^{{\rm H},\infty}({\cal E})\Bigr)^*
\label{mommapdef}\eeq
which naturally generates a system of Hamiltonians $H_{\hat\lambda}:\CC({\cal
E})\to\real$ by
\beq
\left( \mu\left[\hat A\right]\,,\,\hat\lambda\right)=H_{\hat\lambda}
\left[\hat A\right] \ .
\label{Hamdef}\eeq
To determine the moment map explicitly in the present case, we use the
Hamiltonian flow condition
\beq
\frac\delta{\delta\hat A}H_{\hat\lambda}\left[\hat a\right]=-
\omega\left[\delta_{\hat\lambda}\hat A\,,\,\hat a\right] \ , ~~ \hat a\in
\Omega^1({\cal E}) \ ,
\label{Hamflow}\eeq
which is equivalent to the ${\sf G}({\cal E})$-invariance
(\ref{omegainv}). Using (\ref{gaugetransfop}) and (\ref{omegadef}),
the condition (\ref{Hamflow}) reads
\beq
\frac\delta{\delta\hat A}H_{\hat\lambda}\left[\hat a\right]=-
\Tr^{~}_{\cal E}\,\left[\hat\nabla\,,\,\hat\lambda\right]\wedge\hat a \ .
\label{Hamflowexpl}\eeq
Since the trace is invariant under the natural action of the connection on
${\rm End}_{\alg_\theta}({\cal E})$, i.e.
\beq
\Tr^{~}_{\cal E}\left[\hat\nabla_i\,,\,\hat\lambda\right]=0 \ ,
\label{Trintparts}\eeq
using the Leibnitz rule we can write (\ref{Hamflowexpl}) equivalently as
\beq
\frac\delta{\delta\hat A}H_{\hat\lambda}\left[\hat a\right]=\Tr^{~}_{\cal E}\,
\left[\hat\nabla\,\stackrel{\wedge}{,}\,
\hat a\right]\,\hat\lambda \ .
\label{HamflowLeibnitz}\eeq

Let us now compare (\ref{HamflowLeibnitz}) with the first order perturbation of
the shifted field strength in a neighbourhood of a point $\hat A\in\CC({\cal
E})$, which is easily computed to be
\beq
\hat F_{\hat A+t\,\hat a}+\Phi\cdot\id_{\cal E}=\hat F_{\hat A}+\Phi\cdot
\id_{\cal E}+t\left[\hat\nabla\,\stackrel{\wedge}{,}
\,\hat a\right]+O\left(t^2\right) \ .
\label{hatFpert}\eeq
By using (\ref{fnderivdef}) we may thereby
write (\ref{HamflowLeibnitz}) as
\beq
\frac\delta{\delta\hat A}H_{\hat\lambda}\left[\hat a\right]=
\frac\delta{\delta\hat A}
\Tr^{~}_{\cal E}\left(\hat F_{\hat A}+\Phi\cdot\id_{\cal E}\right)\hat
\lambda \ ,
\label{HamflowFA}\eeq
which is equivalent to
\beq
H_{\hat\lambda}\left[\hat A\right]=\left(\hat F_{\hat A}+\Phi\cdot
\id_{\cal E}\,,\,\hat\lambda\right)
\label{Hamfinal}\eeq
in the quadratic form (\ref{Endquad}). Comparing with (\ref{Hamdef}) we see
that the moment map for the action of the noncommutative gauge group on the
space $\CC({\cal E})$ is the shifted noncommutative field strength,
\beq
\mu\left[\hat A\right]=\hat F_{\hat A}+\Phi\cdot\id_{\cal E} \ .
\label{mommapFA}\eeq
Since $\pi^{~}_2\Bigl(U^\infty({\cal E})\Bigr)=0$, the map $\hat\lambda\mapsto
H_{\hat\lambda}$ determines a homomorphism from the Lie algebra ${\rm
End}_{\alg_\theta}^{{\rm H},\infty}({\cal E})$ to the infinite-dimensional
Poisson algebra induced on the space of functions $\CC({\cal E})\to\real$ by
the symplectic two-form (\ref{omegadef}).

\subsection{Cohomological Formulation of Noncommutative Yang-Mills
Theory\label{locNCYM}}

The fact that noncommutative gauge theory is so naturally a Hamiltonian system
leads immediately to the localization of the path integral (\ref{ZNCYMop}) onto
the critical points of the action (\ref{SNCYMop}). We will now sketch the
argument. First of all, the integration measure appearing in (\ref{ZNCYMop}) is
defined to be the Liouville measure corresponding to the symplectic two-form
(\ref{omegadef}),\footnote{\baselineskip=12pt To prove this formula,
  one needs to carefully study the gauge-fixed path integral
  measure. Since the gauge-fixed quantum action in two dimensions is
  Gaussian in the Faddeev-Popov ghost fields, and (\ref{omegadef}) coincides
with
  the symplectic structure of the commutative case, the same arguments
  as in the commutative case~\cite{witten1} apply here and
  (\ref{DDAdef}) is indeed the appropriate gauge-invariant measure to
  use on~$\CC({\cal E})$.}
\beq
\DD\hat A=\dd\hat A~\int\limits_{\Pi\,\Omega^1({\cal E})}\dd\hat\psi~
\e^{-\ii\omega\bigl[\hat\psi\,,\,\hat\psi\bigr]} \ ,
\label{DDAdef}\eeq
where $\dd\hat A$ is the ``ordinary'' Feynman measure which may be defined by
using the identification (\ref{EndMN}) and the operator-field
correspondence as
\beq
\dd\hat A=\prod_{a=1}^N\,\prod_{b=1}^N\,\prod_{x\in{\bf T}^2}\dd A_1^{ab}(x)
{}~\dd A_2^{ab}(x) \ .
\label{ddAdef}\eeq
In (\ref{DDAdef}), $\Pi$ denotes the parity reversion operator, $\hat\psi$ are
the odd generators of functions on the infinite dimensional superspace
\beq
\Pi({\cal E})=\CC({\cal E})\oplus\Pi\,\Omega^1({\cal E}) \ ,
\label{PicalEdef}\eeq
and $\dd\hat A~\dd\hat\psi$ is the corresponding functional Berezin measure.

The result of the previous subsection shows that the noncommutative Yang-Mills
action (\ref{SNCYMop}) is proportional to the square of the moment map, in
the quadratic form (\ref{Endquad}), for the symplectic action of the gauge
group ${\sf G}({\cal E})$ on $\CC({\cal E})$,
\beq
S\left[\hat A\right]=\frac{2\pi^2R^2}{g^2}\,\left(\mu\left[\hat A\right]\,,\,
\mu\left[\hat A\right]\right) \ .
\label{SNCYMmu}\eeq
We can linearize the action (\ref{SNCYMmu}) in $\mu$ via a functional Gaussian
integration over an auxiliary field $\hat\phi\in\Omega^0({\cal E})$, and by
using (\ref{Hamdef}) we can write the partition function (\ref{ZNCYMop}) as
\beq
Z(g^2,\theta,\Phi,{\cal E})= \frac1{
{\rm vol}\,{\sf G}({\cal E})}~\int\limits_{\Omega^0({\cal E})}\dd\hat
\phi~\e^{-\frac1{2\beta}\,\bigl(\hat\phi\,,\,\hat\phi\bigr)}\,
\int\limits_{\Pi({\cal E})}\dd\hat A~\dd\hat\psi~\e^{-\ii\bigl(
\omega[\hat\psi,\hat\psi]-H_{\hat\phi}[\hat A]\bigr)} \ ,
\label{ZNCYMlin}\eeq
with the measure $\dd\hat\phi$ defined analogously to (\ref{ddAdef}). Note that
the operator $\hat\phi$ appears only quadratically in (\ref{ZNCYMlin}) and
thereby essentially corresponds to a commutative field. Because the functional
integration measures in (\ref{ZNCYMlin}) are the same as those which occur in
the corresponding commutative case, the only place that noncommutativity is
present is in the field strength which appears in the moment map
(\ref{mommapFA}). Indeed, it is essentially this feature that leads to the
exact solvability of the model, in parallel with its commutative limit.

The representation (\ref{ZNCYMlin}) of noncommutative gauge theory is the crux
of the matter. Notice first that in the weak coupling limit $g^2=0$
($\beta=\infty$), the noncommutative field $\hat\phi$ appears linearly in
(\ref{ZNCYMlin}), and its integration yields the constraint $\mu\left[\hat
A\right]=0$ which localizes the path integral onto gauge connections $\hat A$
of constant curvature $\hat F_{\hat A}=-\Phi\cdot\id_{\cal E}$. The
partition function in this limit then formally computes the symplectic volume
of the moduli space of constant curvature connections modulo noncommutative
gauge transformations. The key property which enables this localization is that
$\hat A\mapsto\mu\left[\hat A\right]$ is a complete Nicolai map which
trivializes the integration over $\CC({\cal E})$. In this respect, the $g^2=0$
limit of (\ref{ZNCYMlin}) is a topological gauge theory, and indeed it
coincides with a noncommutative version of $BF$ theory in two
dimensions~\cite{blauBF}. What is remarkable
though is that the same Nicolai map appears to trivialize to the full theory
(\ref{ZNCYMlin}) at $g^2\neq0$ to a Gaussian integral over $\CC({\cal E})$.
This works up to the points in $\CC({\cal E})$ where this map has
singularities, which coincide with the solutions of the
classical equations of motion of noncommutative gauge theory. Thus in the
generic case the partition function receives only contributions from the
classical noncommutative gauge field configurations.

To make these arguments precise, we first observe that the integral over the
superspace $\Pi({\cal E})$ in (\ref{ZNCYMlin}) is formally the partition
function of an infinite dimensional statistical mechanics system, and, in the
present situation whereby there is a symplectic group action generated by the
Hamiltonian $H_{\hat\phi}\left[\hat A\right]$, it is known that such integrals
can be typically reduced to finite dimensional integrals, or sums, determined
by the critical points of $H_{\hat\phi}\left[\hat A\right]$~\cite{locbook}. The
main
difference here is that there is no temperature parameter in front of the
Hamiltonian through which to expand, but rather the noncommutative field
$\hat\phi$. The argument for localization can nonetheless be carried through by
adapting the non-Abelian localization principle~\cite{Wittenloc} to
the present noncommutative
setting. This is achieved through a study of the cohomology of the infinite
dimensional operator
\beq
\QQ_{\hat\phi}=\Tr^{~}_{\cal E}\left(\hat\psi_i\,\frac\delta{\delta\hat A_i}+
\left[\hat\nabla_i\,,\,\hat\phi\right]\,\frac\delta{\delta\hat\psi_i}\right)
\label{Qhatphidef}\eeq
which is defined on the space
\beq
\Omega_{{\sf G}({\cal E})}={\rm Sym}\,\Omega^0({\cal E})
\otimes\Bigl(\CC({\cal E})\oplus\Pi\,\Omega({\cal E})\Bigr) \ ,
\label{OmegahatcalG}\eeq
where ${\rm Sym}\,\Omega^0({\cal E})$ is the algebra of
gauge-covariant polynomial functions on ${\rm End}_{\alg_\theta}^{{\rm
H},\infty}({\cal E})$. The linear derivation (\ref{Qhatphidef}) acts on the
basic multiplet $\left(\hat A_i\,,\,\hat\psi_i\,,\,\hat\phi\right)$ of the
noncommutative quantum field theory (\ref{ZNCYMlin}) through the transformation
laws
\bea
\left[\QQ_{\hat\phi}\,,\,\hat A_i\right]&=&\hat\psi_i \ , \nn\\
\left\{\QQ_{\hat\phi}\,,\,\hat\psi_i\right\}&=&\left[\hat\nabla_i\,,\,
\hat\phi\right] \ , \nn\\\left[\QQ_{\hat\phi}\,,\,\hat\phi\right]&=&0 \ ,
\label{Qhatphifieldact}\eea
and its square coincides with the generator of an infinitesimal gauge
transformation with gauge parameter $\hat\phi$,
\beq
\left(\QQ_{\hat\phi}\right)^2=\delta_{\hat\phi} \ .
\label{Qhatphi2}\eeq

The key property of the operator (\ref{Qhatphidef}) is that the Boltzmann
weight over $\Pi({\cal E})$ in (\ref{ZNCYMlin}) is annihilated by it,
\beq
\QQ_{\hat\phi}\left(\Tr^{~}_{\cal E}\,\hat\psi\wedge\hat\psi-H_{\hat\phi}
\left[\hat A\right]\right)=0 \ ,
\label{Boltzmannclosed}\eeq
where we have used the Hamiltonian flow equation (\ref{Hamflowexpl}). Via
integration by parts over the superspace $\Pi({\cal E})$, this implies that the
partition function (\ref{ZNCYMlin}) is unchanged under multiplication of the
Boltzmann factor by $\QQ_{\hat\phi}\alpha$ for any gauge-invariant
$\alpha\in\Omega_{{\sf G}({\cal E})}$, i.e.
\beq
\left(\QQ_{\hat\phi}\right)^2\alpha=0 \ .
\label{Qphialpha0}\eeq
In particular, we may write (\ref{ZNCYMlin}) in the form
\bea
&&Z(g^2,\theta,\Phi,{\cal E})= \frac1{
{\rm vol}\,{\sf G}({\cal E})}~\int\limits_{\Omega^0({\cal E})}\dd\hat
\phi~\e^{-\frac1{2\beta}\,\bigl(\hat\phi\,,\,\hat\phi\bigr)}\,
\int\limits_{\Pi({\cal E})}\dd\hat A~\dd\hat\psi~\e^{-\ii\bigl(
\omega[\hat\psi,\hat\psi]-H_{\hat\phi}[\hat A]-t\,\QQ_{\hat\phi}
\alpha[\hat A,\hat\psi]\bigr)} \ . \nn\\&&
\label{ZNCYMlinpert}\eeq
That the right-hand side of (\ref{ZNCYMlinpert}) is independent of the
parameter $t\in\real$ for gauge
invariant $\alpha$ follows by noting that its derivative
with respect to $t$ vanishes upon integrating by parts over $\Pi({\cal E})$,
and using (\ref{Boltzmannclosed}) and (\ref{Qphialpha0}) along with
the Leibnitz rule for the functional derivative operator
(\ref{Qhatphidef}). This will be true so long as the perturbation by
$\QQ_{\hat\phi}\alpha$ yields an effective action which has a
nondegenerate kinetic energy term, and that it does not allow any new
$\QQ_{\hat\phi}$ fixed points to flow in from infinity in field
space. The $t=0$ limit of (\ref{ZNCYMlinpert}) coincides with the
original partition function of noncommutative gauge theory, while its
$t\to\infty$ limit yields the desired reduction for appropriately chosen
$\alpha$.

At this stage we will choose
\beq
\alpha\left[\hat A\,,\,\hat\psi\right]=4\pi^2R^2\,\Tr^{~}_{\cal E}\,
\hat\psi^i\,\left[\hat\nabla_i\,,\,\mu\left[\hat A\right]\right] \ .
\label{alphachoice}\eeq
Substituting (\ref{alphachoice}) into (\ref{ZNCYMlinpert}) using
(\ref{Qhatphidef}), performing the Gaussian integral over
$\hat\phi\in\Omega^0({\cal E})$, and taking the large $t$ limit, we arrive at
\bea
Z(g^2,\theta,\Phi,{\cal E})&=&\frac1{{\rm vol}\,{\sf G}({\cal E})}~
\int\limits_{\Pi({\cal E})}\dd\hat A~\dd\hat\psi~
\e^{-\Tr^{~}_{\cal E}\bigl(\ii\,\hat\psi\wedge\hat\psi+\frac\beta2
\,\mu[\hat A]^2\bigr)}\nn\\&&\times\,\lim_{t\to\infty}\,\exp\left(
-\frac{(4\pi^2R^2)^3}{2g^2}~t^2\,\Tr^{~}_{\cal E}\left[\hat\nabla^i\,,\,
\left[\hat\nabla_i\,,\,\mu\left[\hat A\right]\right]\right]^2\right)
\nn\\&&\times\,\exp\left\{4\pi^2\ii R^2\,t\,\Tr^{~}_{\cal E}
\left(\mu\left[\hat A\right]
\left[\hat\psi_i\,,\,\hat\psi^i\right]-\left[\hat\nabla_i\,,\,\hat\psi^i
\right]\left[\hat\nabla\,\stackrel{\wedge}{,}
\,\hat\psi\right]\right)\right\} \ , \nn\\&&
\label{ZNCYMlinlarget}\eea
where we have further applied the Leibnitz rule along with (\ref{Trintparts}),
and also dropped overall constants for ease of notation. The $\hat\psi$
integrations in (\ref{ZNCYMlinlarget}) produce polynomial functions of the
parameter $t$, and the $\hat A$ integration is therefore suppressed by the
Gaussian term in $t$ as $t\to\infty$. Nondegeneracy of the quadratic form
(\ref{Endquad}) implies that the functional integral thereby becomes
localized near the solutions of the equation
\beq
\left[\hat\nabla^i\,,\,\left[\hat\nabla_i\,,\,\mu\left[\hat A\right]\right]
\right]=0 \ ,
\label{loceqn}\eeq
and it can be written as a sum over contributions which depend only on local
data near the solutions of (\ref{loceqn}). Along with (\ref{Trintparts}) and
the Leibnitz rule, the equation (\ref{loceqn}) implies
\bea
0&=&\Tr^{~}_{\cal E}\,\mu\left[\hat A\right]\left[\hat\nabla^i\,,\,
\left[\hat\nabla_i\,,\,\mu\left[\hat A\right]\right]\right]^2\nn\\&=&
-\Tr^{~}_{\cal E}\left[\hat\nabla_i\,,\,\mu\left[\hat A\right]\right]^2 \ ,
\label{loceqnimplies}\eea
which again by the non-degeneracy of (\ref{Endquad}) is equivalent to
\beq
\left[\hat\nabla_i\,,\,\mu\left[\hat A\right]\right]=0 \ .
\label{YMeqnsop}\eeq
Since $\mu\left[\hat A\right]=\left[\hat\nabla_1\,,\,\hat\nabla_2\right]$, the
equations (\ref{YMeqnsop}) coincide with the classical equations of motion of
the action (\ref{SNCYMop}), i.e. $\delta S\left[\hat
  A\right]/\delta\hat A=0$. This establishes the localization of the partition
function (\ref{ZNCYMop}) of noncommutative gauge theory in two dimensions onto
the space of solutions of the noncommutative Yang-Mills equations. This space
will be studied in detail in the next section.

Although the above technique leads to a formal proof of the
localization of the partition function onto classical gauge field
configurations, it does not yield any immediate useful information as
to the precise form of the function $W$ in (\ref{ZNCYMcritsum})
encoding the quantum fluctuations about the classical
solutions. The infinite-dimensional determinants that arise from
(\ref{ZNCYMlinlarget}) have very large symmetries and are difficult to
evaluate. The fluctuation determinants $W$ will be determined later on
by another technique. From a
mathematical perspective, the action in (\ref{ZNCYMlin}) over
$\Pi({\cal E})$ is the ${\sf G}({\cal E})$-equivariant extension of
the moment map on
$\CC({\cal E})$, the integration over $\Pi({\cal E})$ defines an equivariant
differential form, and the integral over $\hat\phi\in\Omega^0({\cal E})$
defines equivariant integration of such forms. The operator (\ref{Qhatphidef})
is the Cartan differential for the ${\sf G}({\cal E})$-equivariant
cohomology of $\CC({\cal E})$~\cite{locbook}. The localization may
then also be understood via
a mapping onto a purely cohomological noncommutative gauge theory in the limit
$t\to\infty$. These aspects will not be developed any further here.

\newsection{Classification of Instanton Contributions\label{InstClass}}

In the previous section we proved that the partition function is given by a sum
over contributions localized at the classical solutions of the noncommutative
gauge theory. In this section we will classify the instantons of
two-dimensional gauge theory on the noncommutative torus, and later on
explicitly evaluate their contribution to the partition function. By an
``instanton'' here we mean a solution $A_i=A_i^{\rm cl}$ of the classical
noncommutative field equations
\beq
\partial_iF_A+A_i\star'F_A-F_A\star'A_i=0
\label{NCfieldeqs}\eeq
which is not a gauge transformation of the trivial solution $A_i=0$. Here $F_A$
is the noncommutative field strength (\ref{FAdef}). Note that this definition
also includes the unstable modes. In the commutative case,
instanton contributions have a well-known geometrical classification
based on the fundamental group of the spacetime~\cite{AB}. In the
noncommutative setting, however, the role of homotopy groups is played
by the K-theory of the algebra and one must resort to an algebraic
characterization of the contributing projective modules. For irrational values
of the noncommutativity parameter $\theta$, an elegant classification of the
stationary points of noncommutative Yang-Mills theory has been given
in~\cite{rieffel}. In what follows we shall modify this construction
somewhat to more properly suit our purposes.

\subsection{Heisenberg Modules\label{Heisenberg}}

In order to classify the instanton solutions of gauge theory on the
noncommutative torus, we need to specify the topological structures involved.
This requirement leads us into the explicit classification of the projective
modules over the algebra $\alg_\theta$~\cite{ks,connesbook}. They are
classified by the K-theory group~\cite{pv}
\beq
{\rm
  K}_0(\alg_\theta)=\pi^{~}_1\Bigl(\unitary_\infty(\alg_\theta)\Bigr)
=\zed\oplus\zed \ .
\label{K0Atheta}\eeq
The cohomologically invariant trace $\Tr:\alg_\theta\to\complex$
induces an isomorphism ${\rm
K}_0(\alg_\theta)\to\zed+\zed\,\theta\subset\real$ of ordered groups. To each
pair of integers $(p,q)\in{\rm K}_0(\alg_\theta)$ there corresponds a virtual
projector ${\sf P}_{p,q}$ with $\Tr\otimes\tr^{~}_M\,{\sf P}_{p,q}=p-q\theta$.
However, given a projective module $\cal E$ determined by a Hermitian projector
$\sf P$, positivity of the trace implies
\beq
\dim{\cal E}=\Tr^{~}_{\cal E}\,\id_{\cal E}=\Tr\otimes\tr^{~}_M\,{\sf P}=
\Tr\otimes\tr^{~}_M\,{\sf P}\,{\sf P}^\dag\geq0 \ ,
\label{dimcalEpos}\eeq
and so the stable (rather than virtual) projective modules are classified by
the positive cone of ${\rm K}_0(\alg_\theta)$. Thus to each pair of integers
$(p,q)$ we can associate a {\it Heisenberg module} ${\cal
  E}_{p,q}$~\cite{Rieffel83} of positive
Murray-von~Neumann dimension
\beq
\dim{\cal E}_{p,q}=p-q\theta>0 \ .
\label{dimcalEnm}\eeq
Such pairs of integers parameterize the connected components of the infinite
dimensional manifold ${\rm Gr}^{~}_\theta$ of Hermitian projectors of the
algebra
$\alg_\theta$. In what follows we will be interested in studying the critical
points of the noncommutative Yang-Mills action within a given homotopy class of
${\rm Gr}^{~}_\theta$. The integer
\beq
q=\frac1{2\pi\ii}\,\Tr\otimes\tr^{~}_M\,{\sf P}_{p,q}\left[\hat\partial\,,\,
{\sf P}_{p,q}\right]\wedge\left[\hat\partial\,,\,{\sf P}_{p,q}\right]
\label{Chernnumber}\eeq
is the Chern number (or magnetic flux) of the corresponding gauge
bundle~\cite{connestransl}. In
the case of irrational $\theta$, any finitely generated projective module over
the noncommutative torus is either a free module or it is isomorphic to a
Heisenberg module~\cite{rieffelhigher}. We will view free modules as
special instances of Heisenberg
modules obtained by setting $q=0$. Any two projective modules representing the
same element of K-theory are isomorphic.

The main property of Heisenberg modules that we will exploit in the following
is that they always admit a constant curvature connection $\hat\nabla^{\rm
c}\in\CC_{p,q}=\CC({\cal E}_{p,q})$,
\beq
\left[\hat\nabla_1^{\rm c}\,,\,\hat\nabla_2^{\rm c}
\right]=\ii f\cdot\id_{{\cal E}_{p,q}} \ ,
\label{constcurvcond}\eeq
where $f\in\real$ is a constant. In this subsection we shall set $\Phi=0$, as
the background magnetic flux can be reinstated afterwards by the shift
$f\mapsto f+\Phi$. In the presence of supersymmetry, such a field configuration
gives rise to a BPS state~\cite{cmrMatrix,gme}. It also leads to an
explicit representation of the
Heisenberg module ${\cal E}_{p,q}$ as the separable Hilbert
space~\cite{connesbook}
\beq
{\cal E}_{p,q}=L^2(\real)\otimes\complex^q \ , ~~ q\neq0 \ .
\label{calEnmexpl}\eeq
The Hilbert space $L^2(\real)$ is the Schr\"odinger representation of the
Heisenberg commutation relations (\ref{constcurvcond}). By the
Stone-von~Neumann theorem, it is the unique irreducible representation. The
factor $\complex^q$ defines the $q\times q$ representation of the Weyl-'t~Hooft
algebra in two dimensions,
\beq
\Gamma_1\,\Gamma_2=\e^{2\pi\ii p/q}\,\Gamma_2\,\Gamma_1 \ ,
\label{WeyltHooftalg}\eeq
which may be solved explicitly by $SU(q)$ shift and clock matrices. The
generators of the noncommutative torus are then represented on
(\ref{calEnmexpl}) as
\beq
\hat Z_i=\e^{\ii f^{-1}\,\hat\nabla_i^{\rm c}/R}\otimes\Gamma_i \ ,
\label{hatZirep}\eeq
and computing (\ref{NCcommrel}) using (\ref{constcurvcond}),
(\ref{WeyltHooftalg}) and the Baker-Campbell-Hausdorff formula thereby leads to
a relation between the noncommutativity parameter $\theta$ and the constant
flux $f$ through
\beq
\theta=-\frac1{2\pi R^2f}+\frac pq \ .
\label{thetafrel}\eeq
The $\alg_\theta$-valued inner product on ${\cal E}_{p,q}$ is given by
\bea
\left\langle\hat\xi\,,\,\hat\eta\right\rangle_{\alg_\theta}^{~}&=&
\sum_{m_1=-\infty}^\infty~\sum_{m_2=-\infty}^\infty~\left(~\int
\limits_{-\infty}^\infty\dd s~\left[\Gamma_1^{m_1}\,\Gamma_2^{m_2}\,
\xi\left(s-\frac{m_1}{2\pi R^2f}\right)\right]^\dag\,\eta(s)~
\e^{2\pi\ii m_2}\right)\nn\\&&\times\,\hat Z_1^{m_1}\hat Z_2^{m_2} \ .
\label{Epqinnerprod}\eea
For $q=0$ we define ${\cal E}_{p,0}$ to be the free module of rank $p$, i.e.
\beq
{\cal E}_{p,0}=L^2({\bf T}^2)\otimes\complex^p \ .
\label{calEfree}\eeq

The Heisenberg module ${\cal E}_{p,q}$ so constructed coincides, in the D-brane
picture, with the Hilbert space of ground states of open strings stretching
between a single D$r$-brane and $p$ D$r$-branes carrying $q$ units of
D$(r-2)$-brane charge~\cite{sw}. It is irreducible if and only if the
integers $p$ and $q$ are relatively
prime. The Weyl-'t~Hooft algebra (\ref{WeyltHooftalg}) has a unique irreducible
representation (up to $SU(q)$ equivalence) of dimension $q/{\rm
  gcd}(p,q)$~\cite{vanBaal,lebedev}, and
so the rank $N$ of the resulting gauge theory as defined in
section~\ref{nctorusym} is given by
\beq
N={\rm gcd}(p,q) \ .
\label{Ngcdnm}\eeq
Furthermore, the commutant $\mat_N(\alg_{\theta'})$ of $\alg_\theta$ in ${\rm
  End}_{\alg_\theta}({\cal E}_{p,q})$ is Morita equivalent to the
  noncommutative torus with dual noncommutativity parameter $\theta'$
  determined by the $SL(2,\zed)$ transformation~\cite{connesbook}
\beq
\theta'=\frac{n-s\theta}{p-q\theta}\,N \ ,
\label{thetaprimeexpl}\eeq
where $n$ and $s$ are integers which solve the Diophantine equation
\beq
ps-qn=N \ .
\label{Dioeq}\eeq

\subsection{Stationary Points of Noncommutative Gauge Theory\label{Critical}}

We will now describe the critical points of the noncommutative Yang-Mills
action (\ref{SNCYMop}). Let us fix a Heisenberg module ${\cal E}_{p,q}$ over
the noncommutative torus, which is labelled by a pair of integers $(p,q)$
obeying
the constraint (\ref{dimcalEnm}). From (\ref{constcurvcond}) and
(\ref{thetafrel}) it follows that this projective module is characterized by a
connection $\hat\nabla^{\rm c}\in\CC_{p,q}$ of constant curvature
\beq
\hat F_{\hat A^{\rm c}}=\frac1{2\pi R^2}\,\frac{q}{p - q \theta} \cdot
\id_{ {\cal E}_{p,q}} \ .
\label{constcurvcondF}\eeq
Such constant curvature connections are of fundamental importance in finding
solutions of the noncommutative Yang-Mills equations because they not only
solve (\ref{NCfieldeqs}), but they moreover yield the absolute minimum value of
the Yang-Mills action on the module ${\cal E}_{p,q}$~\cite{amns,cr}. This
follows by using (\ref{hatFpert}) to compute the infinitesimal variation $\hat
F_{\hat A^{\rm c}+t\,\hat a}$ about a constant curvature connection to get
\bea
S\left[\hat\nabla^{\rm c}+t\,\hat{a}\right]&=&\frac{2\pi^2R^2}{g^2}\,
\Tr^{~}_{{\cal E}_{p,q}}\left(\hat{F}_{\hat A^{\rm c} +t\,\hat{a} } + \Phi
\cdot
\id_{{\cal E}_{p,q}} \right)^2\nn\\&=&S\left[\hat\nabla^{\rm c}\right]
+\frac{2\pi^2R^2\,t^2}{g^2}\,\Tr^{~}_{{\cal E}_{p,q}}
\left[\hat\nabla^{\rm c}\,\stackrel{\wedge}{,}
\,\hat a\right]^2+O\left(t^4\right) \ .
\label{SNCYMpert}\eea
The cross terms of order $t$ in (\ref{SNCYMpert}) vanish due to the
property (\ref{Trintparts}) and the fact that the field strength
(\ref{constcurvcondF}) is proportional to the identity operator on ${\cal
E}_{p,q}$. Since the quadratic term in $t$ is positive definite, we have
$S\left[\hat\nabla^{\rm c}+ \hat{a}\right]\geq S\left[\hat\nabla^{\rm
c}\right]~~\forall\hat a\in\Omega^1({\cal E}_{p,q})$. To establish
that $\hat\nabla^{\rm c}$ is a {\it global} minimum, we can exploit
the freedom of choice of the background flux $\Phi$ (see section~6.1)
to identify it with the constant curvature
(\ref{constcurvcondF}). Then $S\left[\hat\nabla^{\rm c}\right]=0$, and
since (\ref{SNCYMop}) is a positive functional, the claimed property
follows. This shifting of the curvature will be used explicitly below.

In addition to yielding the minimum of the Yang-Mills action, constant
curvature connections can also be used to construct {\it all} solutions of the
classical equations of motion~\cite{rieffel}. The main observation is that
insofar as solutions of the Yang-Mills equations are concerned, the module
${\cal E}_{p,q}$ may be considered to be a direct sum of submodules~\cite{AB}.
To see this, we note that the equations of motion (\ref{YMeqnsop}) imply that,
at the critical points $\hat\nabla=\hat\nabla^{\rm cl}$, the moment map
$\mu\left[\hat A\right]$ is invariant under the induced action of the
Heisenberg
algebra ${\cal L}_\Phi$ of automorphisms on the algebra ${\rm
End}_{\alg_\theta}({\cal
E}_{p,q})$. In particular, it corresponds to the central element of the
Heisenberg
Lie algebra generated by $\hat\nabla_1^{\rm cl}$, $\hat\nabla_2^{\rm cl}$ and
$\mu\left[\hat A^{\rm cl}\right]$. This feature provides a natural direct sum
decomposition of the module ${\cal E}_{p,q}$ through the adjoint action of the
moment map on $\Omega_{p,q}=\Omega({\cal E}_{p,q})$. For this, we consider the
self-adjoint linear operators $\Xi_{\hat\nabla}:\Omega_{p,q}\to\Omega_{p,q}$
defined for each connection $\hat\nabla\in\CC_{p,q}$ by
\beq
\Xi_{\hat\nabla}
(\hat\alpha)=\left[\mu\left[\hat A\right]\,,\,\hat\alpha\right] \ ,
{}~~ \hat\alpha\in\Omega_{p,q} \ .
\eeq
{}From the equations of motion (\ref{YMeqnsop}) it follows that the
$\alg_\theta$-valued eigenvalues $\hat c_k$ of $\Xi_{\hat\nabla}$ are constant
in the vicinity of a critical point $\hat\nabla=\hat\nabla^{\rm cl}$, and so
there is a natural direct sum decomposition of the module ${\cal E}_{p,q}$ into
projective submodules ${\cal E}_{p_k,q_k}$,
\beq
{\cal E}_{p,q} = \bigoplus_{k\geq1}{\cal E}_{p_k, q_k} \ ,
\label{calEpqdecomp}\eeq
corresponding to the eigenspace decomposition
$\Omega_{p,q}=\bigoplus_{k\geq1}\Omega_{p_k,q_k}$ with respect to
$\Xi_{\hat\nabla}$.\footnote{\baselineskip=12pt It should be stressed
  that (\ref{calEpqdecomp}) is {\it not} the statement that the given
  Heisenberg module is reducible. It simply reflects the behaviour of
  connections near a stationary point of the noncommutative Yang-Mills
  action, in which one may interpret the eigenspaces
  $\Omega_{p_k,q_k}=\Omega({\cal E}_{p_k,q_k})$ as the differential
  algebras of submodules ${\cal E}_{p_k,q_k}\subset{\cal
    E}_{p,q}$. For more technical details of the decomposition
  (\ref{calEpqdecomp}) as an $\alg_\theta$-module, we refer
  to~\cite{rieffel}. Notice also that here we abuse notation by making
  no distinction between $\hat\nabla^{\rm c}$ acting on $\Omega_{p,q}$
  or ${\cal E}_{p,q}$. Only the latter operator will be pertinent in
  what follows.} On each $\Omega_{p_k,q_k}$ the operator $\Xi_{\hat\nabla}$
acts as multiplication by a fixed scalar $c_k$. Since $\mu\left[\hat A^{\rm
cl}\right]$ commutes with $\hat\nabla^{\rm cl}$, the connection
$\hat\nabla^{\rm cl}$ is also a linear operator on each ${\cal
E}_{p_k,q_k}\to{\cal E}_{p_k,q_k}$, and its restriction $\hat\nabla^{\rm
c}_{(k)}=\left.\hat\nabla^{\rm cl}\right|_{{\cal E}_{p_k,q_k}}$ has constant
curvature $\left.\mu\left[\hat A^{\rm cl}\right]\right|_{{\cal E}_{p_k,q_k}}$.

Given such a direct sum decomposition\footnote{\baselineskip=12pt In
section~\ref{INCYM} we will give an elementary proof that such decompositions
necessarily contain only a finite number of direct
summands. See~\cite{rieffel} for a functional analytic proof.} of the
module ${\cal E}_{p,q}$, we can define a connection $\hat\nabla$ on
${\cal E}_{p,q}$ by taking the sum of connections on each of the
submodules, $\hat{\nabla}=
\bigoplus_{k\geq1}\hat{\nabla}^{~}_{(k)}$. The noncommutative Yang-Mills action
is additive with respect to this decomposition,
\beq
S\left[\,\bigoplus_{k\geq1}\hat{\nabla}^{~}_{(k)}\right]=
\sum_{k\geq1}S\left[\hat{\nabla}^{~}_{(k)}\right] \ .
\label{SNCYMadd}\eeq
It follows that for the particular choice of constant curvature connections
$\hat\nabla_{(k)}^{\rm c}$ on each of the submodules ${\cal E}_{p_k,q_k}$, the
Yang-Mills action has a critical point
\beq
\hat\nabla^{\rm cl}=\bigoplus_{k\geq1}\hat\nabla_{(k)}^{\rm c}
\label{nablacritpoint}\eeq
on ${\cal E}_{p,q}$. Moreover, from the above arguments it also follows that
every Yang-Mills critical point on ${\cal E}_{p,q}$ is of this form.

This construction thereby exhausts all possible critical points, and is
essentially the noncommutative version of the bundle splitting method of
constructing classical solutions to ordinary, commutative gauge theory
in two dimensions~\cite{AB}. While there are many possibilities for
the decomposition (\ref{calEpqdecomp}) of the given module ${\cal
  E}_{p,q}$ into submodules, there are two important constraints that
must be taken into account. First of all, the (positive)
Murray-von~Neumann dimension of the module is additive with respect to
the decomposition~(\ref{calEpqdecomp}),
\beq
\dim{\cal E}_{p,q}=\sum_{k\geq1}\dim{\cal E}_{p_k,q_k}
=\sum_{k\geq1}\left(p_k - q_k \theta\right) \ .
\label{additup}\eeq
Secondly, since a module over the noncommutative torus is completely and
uniquely determined (up to isomorphism) by two integers, we need an additional
constraint. This is the requirement that the Chern number of the module be
equal to the total magnetic flux of the direct sum decomposition. For the
module ${\cal E}_{p,q}$ this gives the relation
\beq
q = \sum_{k\geq1} q_k \ .
\label{chernconstraint}\eeq
For irrational values of the noncommutativity parameter $\theta$ it is clear
from (\ref{dimcalEnm}) that the constraint (\ref{chernconstraint}) follows from
(\ref{additup}). This is not the case for rational $\theta$ and the constraint
on the Chern class makes necessary a distinction between physical
noncommutative Yang-Mills theory and Yang-Mills theory defined on a particular
projective module ${\cal E}_{p,q}$. The latter field theory imposes a
K-theory charge conservation law for submodule decompositions
(\ref{calEpqdecomp}), $(p,q)=\sum_{k\geq1}(p_k,q_k)$. This distinction
will be discussed further
when the partition function for Yang-Mills theory on the noncommutative torus
is calculated explicitly.

We can now summarize the classification of the critical points of the
noncommutative Yang-Mills action as follows. For any value of the
noncommutativity parameter $\theta$, any solution of the classical equations of
motion of Yang-Mills theory defined on the Heisenberg module ${\cal E}_{p,q}$
is completely characterized by a collection of pairs of integers
$\Bigl\{(p_k,q_k)\Bigr\}_{k\geq1}$ obeying the constraints
\bea
p_k -  q_k\theta&>&0 \ , \nn\\
\sum_{k\geq1}\left(p_k - q_k \theta\right)&=&p - q \theta \ , \nn \\
\sum_{k\geq1} q_k&=&q \ .
\label{partrules}
\eea
We will call such a collection of integers a ``partition'' \footnote{
\baselineskip=12pt This definition of partition is more
general than that of~\cite{rieffel}. It is the one that is the most
useful for the computation of the
noncommutative gauge theory partition function in the following. In
particular, it contains contributions from reducible connections, as
these will also turn out to contribute to the Yang-Mills partition
function. These connections are in fact responsible for the orbifold
singularities that appear in the instanton moduli spaces. These
points, as well as how to avoid the overcounting of critical points
through combinatoric factors in the partition function, will be
described in detail in section~9.} and will denote it by
$(\bfp,\bfq)\equiv\Bigl\{(p_k,q_k)\Bigr\}_{k\geq1}$. In order to avoid
overcounting partitions which will contribute to the Yang-Mills
partition function, we also need to introduce a partial ordering for
submodules in a given partition based on the dimension of each submodule,
\beq
0< p_1 - q_1 \theta \leq p_2 - q_2 \theta \leq p_3 -q_3 \theta\leq\dots \ .
\eeq
Any number of partitions which are identical after such an ordering will be
regarded as equivalent presentations of the same partition.
The set of all distinct
partitions associated with the Heisenberg module
${\cal E}_{p,q}$ will be denoted ${\cal P}_{p,q}(\theta)={\cal P}
(\theta,{\cal E}_{p,q})$.

It remains to evaluate the Yang-Mills action at a solution of the classical
equations of motion, which, by the arguments of the previous section, is one of
the key ingredients in the computation of the partition function of
noncommutative gauge theory. At a critical point, i.e. a partition
$(\bfp,\bfq)$, according to (\ref{SNCYMadd}) it is just the sum of
contributions from constant curvature connections on each of the submodules of
the partition,
\beq
S(\bfp,\bfq;\theta) =S\left[\,\bigoplus_{k\geq1}\hat\nabla_{(k)}^{\rm c}
\right]=\frac1{2g^2R^2}\,
\sum_{k\geq1}\left(p_k - q_k\theta\right)\left( \frac{ q_k}{p_k - q_k\theta}
- \frac{q}{p - q\theta} \right)^2 \ ,
\label{genymaction}
\eeq
where we have used $\Tr^{~}_{{\cal E}_{p_k,q_k}}\id_{{\cal
E}_{p_k,q_k}}=p_k-q_k\theta$ and fixed the value of the background field $\Phi$
to be~\cite{amns}
\beq
\Phi = \Phi_{p,q}=- \frac{q}{2R^2( p- q \theta)} \ .
\label{Phipq}\eeq
This value ensures that the constant curvature connection $\hat\nabla^{\rm c}$
on the module ${\cal E}_{p,q}$, which corresponds to a solution of the
equations of motion parameterized by the trivial partition $(p,q)$, gives a
vanishing global minimum of the action, $S(p,q;\theta)=0$. This is the natural
boundary condition, and generally the inclusion of $\Phi$ ensures that
the classical action is invariant under Morita
duality~\cite{amns,sw,gme,schwarzPhi,Seiberg}.

\newsection{Yang-Mills Theory on a Commutative Torus\label{CYM}}

As we mentioned in section~\ref{partfnloc}, while we can prove that
noncommutative gauge theory on a two-dimensional torus is given exactly by a
sum over classical solutions (instantons), evaluating directly the fluctuation
factors, which multiply the Boltzmann weights of the corresponding critical
action values computed in the previous section, is a difficult task. We will
therefore proceed as follows. We start with the well-known exact solution for
Yang-Mills theory on a {\it commutative} torus and identify quantities which
are invariant under gauge Morita equivalence. This will yield the partition
function of noncommutative Yang-Mills theory for any {\it rational} value of
the noncommutativity parameter $\theta$. From this expression we will then be
able to deduce the corresponding expression for Yang-Mills theory defined on a
noncommutative torus with arbitrary $\theta$. In this section we will analyze
the instanton contributions to commutative Yang-Mills theory in order
to set up this construction.

The physical Hilbert space ${\cal H}_{\rm phys}$ of ordinary $U(p)$
quantum gauge theory defined on a
(commutative) two-torus is the space of class functions
\beq
{\cal H}_{\rm phys}=L^2\Bigl(U(p)\Bigr)^{{\rm Ad}\bigl(U(p)\bigr)}
\label{Hilbertphys}\eeq
in the invariant Haar measure on the $U(p)$ gauge group. By the
Peter-Weyl theorem, it has a natural basis
$|{\cal R}\rangle$ determined by the unitary irreducible representations $\cal
R$ of the unitary Lie group $U(p)$. The Hamiltonian is essentially the
Laplacian on the group manifold of $U(p)$, and so the corresponding vacuum
amplitude has the well-known heat kernel
expansion~\cite{witten1,migdal,rusakov}
\beq
Z(g^2,p)=\sum_{\cal R}\e^{-2\pi^2R^2g^2\,C_2({\cal R})} \ ,
\label{repcomYM}\eeq
where the Boltzmann weight contains the quadratic Casimir invariant $C_2({\cal
R})$ of the representation $\cal R$. This concise form does not have a direct
interpretation in terms of a sum over contributions from critical points of the
classical action that we expect from the arguments of section~\ref{partfnloc}.
In order to find a more appropriate form, it is useful to make explicit the sum
over irreducible representations as a sum over integers and perform a Poisson
resummation of (\ref{repcomYM})~\cite{griguolo1}.

The representations $\cal R$ of $U(p)$ can be labelled by sets of $p$ integers
\beq
+\infty>n_1>n_2>\dots>n_p>-\infty
\label{Rints}\eeq
which give the lengths of the rows of the corresponding Young
tableaux. In terms of these integers the Casimir operator is given by
\beq
C_2({\cal R})=C_2(n_1,\dots,n_p)=\frac p{12}\,\Bigl(p^2-1\Bigr)+\sum_{a=1}^p
\left(n_a-\frac{p-1}2\right)^2 \ ,
\label{C2nk}\eeq
and by using its symmetry under permutations of the integers $n_a$ we can write
(\ref{repcomYM}) as
\beq
Z(g^2,p)=\frac1{p!}\,
\sum_{n_1\neq\dots\neq n_p}\e^{-2\pi^2R^2g^2\,C_2(n_1,\dots,n_p)} \ .
\label{comYMints}\eeq
One can extend the sums in (\ref{comYMints}) over {\it all} integers
$n_1,\dots,n_p$ by inserting the determinant
\beq
\det_{1\leq a,b\leq p}\,\left(\delta_{n_a,n_b}\right)=\sum_{\sigma\in\Sigma_p}
{\rm sgn}(\sigma)\,\prod_{a=1}^p\delta_{n_a,n_{\sigma(a)}} \ ,
\label{detdelta}\eeq
where $\Sigma_p$ is the group of permutations on $p$ objects. The permutation
symmetry of (\ref{C2nk}) implies that all elements in the same conjugacy class
of $\Sigma_p$ yield the same contribution to the partition function. The sum
over permutations (\ref{detdelta}) thereby truncates to a sum over conjugacy
classes of $\Sigma_p$. They are labelled by the sets of $p$ integers
$0\leq\nu_a\leq[p/a]$, each giving the number of elementary cycles of length
$a$ in the usual cycle decomposition of elements of $\Sigma_p$, and
which define a partition of $p$, i.e.
\beq
\nu_1+2\nu_2+\dots+p\nu_p=p \ .
\label{conjclass}\eeq
The parity of the elements of a conjugacy class
$C[\vec\nu\,]=C[\nu_1,\dots,\nu_p]$ is
\beq
{\rm sgn}\,C[\vec\nu\,]=(-1)^{\sum_{a'}\nu_{2a'}}
\label{conjparity}\eeq
and it contains
\beq
\Bigl|C[\vec\nu\,]\Bigr|=\frac{p!}{\displaystyle\prod_{a=1}^pa^{\nu_a}\,\nu_a!}
\label{conjnumber}\eeq
elements.

The sum over the $n_a$'s in (\ref{comYMints}) then yields a theta-function, and
the corresponding Jacobi inversion formula can be derived in the usual way by
means of the Poisson resummation formula
\beq
\sum_{n=-\infty}^\infty f(n)=\sum_{q=-\infty}^\infty~\int
\limits_{-\infty}^\infty\dd s~f(s)~\e^{2\pi\ii qs} \ .
\label{Poissonresum}\eeq
The Fourier transformations required in (\ref{Poissonresum}) are all Gaussian
integrals in the present case, and after some algebra the partition function
(\ref{comYMints}) can be expressed as a sum over dual integers $q_k$
as~\cite{griguolo1}~\footnote{\baselineskip=12pt We have corrected here a few
typographical errors appearing in eqs.~(13)--(15) of~\cite{griguolo1}.}
\bea
Z(g^2,p)&=&\e^{-\frac{\pi^2g^2R^2}6\,p(p^2-1)}\,
\sum_{\vec\nu\,\,:\,\sum_aa\nu_a=p}~
\sum_{q_1=-\infty}^\infty\cdots\sum_{q_{|\vec\nu\,|}
=-\infty}^\infty\e^{\ii\pi\bigl(
\sum_{a'}\nu_{2a'}+(p-1)\sum_kq_k\bigr)}\nn\\&&\times\,
\prod_{a=1}^p\frac{\left(2g^2R^2a^3\right)^{-\nu_a/2}}{\nu_a!}~
\e^{ - S_{\vec\nu\,}(q_1,\ldots, q_{|\vec\nu\,|})} \ .
\label{comym}\eea
Here
\beq
|\vec\nu\,|=\nu_1+\nu_2+\dots+\nu_p
\label{nutotal}\eeq
is the total number of cycles contained in the elements of the conjugacy class
$C[\vec\nu\,]$ of $\Sigma_p$, and the action is given by
\bea
S_{\vec\nu\,}(q_1,\ldots,q_{|\vec\nu\,|})&=&\frac1{2g^2R^2}\,
\left(\,\sum_{k_1=1}^{\nu_1} \frac{q_{k_1}^2}{1} +
\sum_{k_2=\nu_1+1}^{\nu_1+\nu_2}
\frac{q_{k_2}^2}{2} + \sum_{k_3=\nu_1+\nu_2+1}^{\nu_1+\nu_2+\nu_3}
\frac{q_{k_3}^2}{3}\right.\nn\\&&+\left.\dots+
\sum_{k_p=\nu_1 +\dots+\nu_{p-1}+1}^{|\vec\nu\,|} \frac{q_{k_p}^2}{p}\right) \
{}.
\label{comymaction}\eeq
It is understood here that if some $\nu_a=0$, then
$q_{\nu_1+\dots+\nu_{a-1}+1}=\dots=q_{\nu_1+\dots\nu_{a-1}+\nu_a}=0$.

The remarkable feature of this rewriting is that the action (\ref{comymaction})
is precisely of the general form (\ref{genymaction}). Since ${\rm
K}_0\Bigl(C^\infty({\bf T}^2)\Bigr)=\zed\oplus\zed$, any
finitely-generated projective module ${\cal E}={\cal E}_{p,q}$ over
the algebra $\alg_0=C^\infty({\bf T}^2)$ is determined (up to
isomorphism) by a pair of relatively prime integers
$(p,q)\in\zed_+\oplus\zed$ with dimension given by $p$ and constant
curvature $q/p$~\cite{Rieffel83}. Geometrically, ${\cal E}_{p,q}$ is
the space of sections of a vector bundle over the torus ${\bf T}^2$ of
rank $p$, topological charge $q$, and with structure group
$U(p)$. Consider a direct sum decomposition (\ref{calEpqdecomp}) of
this module. We will enumerate submodules
in a partition according to increasing dimension. Let $\nu_a$ be the number of
submodules of dimension $a$, corresponding to the splitting of the bundle into
sub-bundles of rank $a$, so that
\beq
\dim{\cal E}_{p,q}=\nu_1 + 2  \nu_2  + \dots + p \nu_p \ .
\eeq
This condition is simply the constraint (\ref{additup}) on the total dimension
of the sum of submodules in this case, i.e. $p=\sum_{k\geq1}p_k$ with $1\leq
p_k\leq p$. Therefore, the expression (\ref{comym}) is nothing but the
localization of the partition function of commutative Yang-Mills theory onto
its classical solutions. Note that here the magnetic charges $q_k$ are dual to
the lengths of the rows of the Young tableaux of the unitary group $U(p)$.

There are, however, two important differences here. First of all, the action
(\ref{comymaction}) is evaluated for a topologically trivial bundle, i.e.
$q=0$, which yields a vanishing background flux $\Phi_{p,q}$. Consequently,
(\ref{comym}) is not the most general result. Secondly, and most importantly,
the sum over Chern numbers $q_1,\dots,q_{|\vec\nu\,|}$ in the
partition function is not constrained to satisfy
(\ref{chernconstraint}), which in view of our first point is the
restriction $\sum_kq_k=0$. In fact, the partition function
(\ref{comym}) for {\it physical} $U(p)$ Yang-Mills gauge theory on the
commutative torus is a sum of contributions from topologically distinct bundles
(of different Chern numbers) over the torus.  In order to generalize the
calculation of the partition
function to the case of Yang-Mills theory defined on a projective
module, we need to separate out of (\ref{comym}) the terms which
are well-defined on a particular isomorphism class ${\cal E}_{p,q}$ of modules.

In order to facilitate the identification of such a module definition of
Yang-Mills theory, we write the partition function (\ref{comym}) in terms of
the topological numbers of the module ${\cal E}_{p,q}$.  We will first enforce
the constraint (\ref{chernconstraint}) on the magnetic charges. It is also
useful
for further calculations to re-interpret the parity factors
$(-1)^{\sum_{a'}\nu_{2a'} }$ in terms of the rank $p$ and the total number
$|\vec\nu\,|$ of submodules in a given partition
$(\bfp,\bfq)=\Bigl\{(p_k,q_k)\Bigr\}_{k=1}^{|\vec\nu\,|}$ labelling a
critical point of the action. If $p$ is odd (even) then there is an
odd (even) number of
submodules ${\cal E}_{p_k,q_k}$ with $p_k$ odd. By considering all possible
cases one can show that
\beq
\sum_{a'=1}^{[p/2]}\nu_{2a'}=p +|\vec\nu\,|~~({\rm mod}~2) \ .
\eeq
With these adjustments we are led to the module Yang-Mills theory
with partition function $Z_{p,q}$ which is well-defined on ${\cal E}_{p,q}$,
\beq
Z(g^2,p)= \e^{- \frac{\pi^2g^2R^2}6\,p (p^2 -1) }\,\sum_{q=-\infty}^\infty
(-1)^{ (p-1)q +p}~Z_{p,q}(g^2,\theta=0) \ ,
\label{partfnphys}\eeq
where the module partition function is given by a sum over partitions
associated with the module ${\cal E}_{p,q}$,
\bea
Z_{p,q}(g^2,\theta=0)&=&Z(g^2,\theta=0,\Phi_{p,q},{\cal E}_{p,q})\nn\\
&=&\sum_{(\bfp,\bfq)\in{\cal P}_{p,q}(\theta=0)}(-1)^{|\vec\nu\,|}\,
\prod_{a=1}^p\frac{\left(2g^2R^2a^3\right)^{-\nu_a/2}}{\nu_a!}
{}~\e^{ - S(\bfp,\bfq;\theta=0)} \ .
\label{commodYM}\eeq
Note that the critical points of the action  are defined by partitions
obeying the constraints (\ref{partrules}), including a restriction to
submodules with total Chern number $q$.  We have also generalized
to the correct action (\ref{genymaction}) for Yang-Mills theory on a bundle
with Chern number $q$ which contains the non-vanishing value (\ref{Phipq}) for
the background magnetic field $\Phi$.  Again, this latter change is equivalent
to adding boundary terms to the action which do not contribute to the
classical dynamics
of the theory and hence are not relevant to our analysis based on instanton
contributions. The only essential role of $\Phi$ is, as we will see,
to set the zero-point of the Yang-Mills action in the instanton
picture. Therefore, a shift in $\Phi$ will at most result in
multiplying the fixed module partition function (\ref{commodYM}) by
overall constants dependent only on the topological numbers $(p,q)$.

\newsection{Yang-Mills Theory on a Noncommutative Torus: Rational
Case\label{RNCYM}}

Given the partition function (\ref{commodYM}) for Yang-Mills theory which is
well-defined on a given module ${\cal E}_{p,q}$ of sections of some bundle, we
can now use Morita equivalence to obtain an explicit formula for Yang-Mills
theory on a torus with rational noncommutativity parameter $\theta$ from the
commutative case. Morita equivalence in this case refers to the mapping between
noncommutative tori which is generated by the infinite discrete group
$SO(2,2,\zed)\cong SL(2,\zed)\times SL(2,\zed)$, where one of the $SL(2,\zed)$
factors coincides with the discrete automorphism group of the ordinary torus
${\bf T}^2$. It provides a one-to-one correspondence between modules associated
with different topological numbers and noncommutativity parameters. Here we
will be interested in the transformations from modules corresponding to
rational values of $\theta$ to modules with vanishing $\theta$. In fact, there
is an extended version of the correspondence known as gauge Morita
equivalence~\cite{sw,gme} which augments the mapping of tori with
transformations of connections between modules, and leads to a rescaling of the
area and coupling constant to give a symmetry of Yang-Mills theory as we have
defined it in (\ref{ZNCYMfield}). The entire noncommutative quantum field
theory is invariant under this extended equivalence~\cite{amns} which coincides
with the standard open string T-duality
transformations~\cite{sw,schwarzPhi}. We will use this
invariance property to construct the noncommutative gauge theory for rational
values of the deformation parameter~$\theta$.

\subsection{Gauge Morita Equivalence\label{Morita}}

We begin by summarizing the basic transformation rules of Morita equivalence of
noncommutative gauge theories~\cite{ks,sz1,Rieffel83}. In two
dimensions, Morita equivalences of
noncommutative tori are generated by the group elements
\beq
\pmatrix{m&n\cr r&s\cr}\in SL(2,\zed) \ ,
\label{Moritamatrix}\eeq
where we concentrate on the $SL(2,\zed)$ subgroup which acts only on the
K\"ahler modulus of ${\bf T}^2$. The full duality group acts on the
K-theory ring ${\rm K}_0(\alg_\theta)\oplus{\rm K}_1(\alg_\theta)$ in a spinor
representation of $SO(2,2,\zed)$ and the topological numbers $(p,q)$
of a module ${\cal E}={\cal E}_{p,q}$ transform as
\beq
\pmatrix{p'\cr q'\cr}=\pmatrix{m&n\cr r&s\cr}\pmatrix{p\cr q\cr} \ .
\label{pqspinor}\eeq
The noncommutativity parameter $\theta$ transforms under a discrete linear
fractional transformation
\beq
\theta^\prime=\frac{m\theta+n}{r\theta+s} \ .
\eeq
{}From these rules it follows that under the gauge Morita equivalence
parameterized by (\ref{Moritamatrix}) the dimensions of modules are changed
according to
\beq
\dim {\cal E}^\prime =\frac{\dim {\cal E}}{|r\theta+s|} \ .
\label{dimEprime}\eeq
The invariance of the noncommutative Yang-Mills action (\ref{SNCYMfield}) then
dictates the corresponding transformation rules for the area element of ${\bf
T}^2$, the Yang-Mills coupling constant, and the magnetic background as
\bea
R'&=&|r\theta+s|\,R \ , \nn\\g'^2&=&|r\theta+s|\,g^2 \ , \nn\\
\Phi'&=&(r\theta+s)^2\,\Phi-\frac{r(r\theta+s)}{2\pi R^2} \ .
\label{othertransfs}\eeq

\subsection{The Partition Function for Rational $\theta$\label{RatPart}}

Let us now consider the effect of such transformations on the module partition
function defined in (\ref{commodYM}). Under the gauge Morita equivalence
parameterized by (\ref{Moritamatrix}), the ordinary Yang-Mills gauge theory
(\ref{commodYM}) is mapped onto a noncommutative gauge theory with
rational-valued noncommutativity parameter $\theta=n/s$. The classical action
of the theory is invariant, and constant curvature connections are mapped into
one another~\cite{ks,gme}. Thus Morita equivalence maps solutions of
the equations of motion
between the commutative and rational noncommutative cases. The localization of
the partition function onto classical solutions is therefore not affected by
the transformation. The topological numbers of the submodules comprising
partitions which define solutions of the classical equations of motion also map
into each other in the two cases. In particular, the total number
$|\vec\nu\,|$ of submodules in a partition is invariant under the
Morita duality.

The only component of the partition function (\ref{commodYM}) we have left to
examine is the pre-exponential factor
$\prod_{a\geq1}\,(2g^2R^2a^3)^{-\nu_a/2}/\nu_a!$. The symmetry factors
$\nu_a!$ associated with a partition are preserved, and so from the
transformation rules
(\ref{othertransfs}) for $\theta=0$ it follows that this component is invariant
only if the integer $a^3$ appearing here transforms according to the scaling
\beq
a'=\frac a{|s|}
\label{kprime}\eeq
under the Morita equivalence. But (\ref{kprime}) is exactly the rescaling
(\ref{dimEprime}) of the dimension of a projective module in this case. It
follows that the indices $a$ in the pre-exponential factors of (\ref{commodYM})
should be interpreted as the (integer) dimensions of submodules in the
commutative gauge theory, and this fact provides a Morita covariant
interpretation of these indices which leads immediately to the appropriate
generalization of the formula (\ref{commodYM}) to rational-valued
$\theta\neq0$.

We are now in a position to write down an explicit expression for the partition
function of quantum Yang-Mills theory on the module ${\cal E}_{p,q}$
corresponding to rational, non-integer noncommutativity parameter
$\theta$. The only modifications required are the counting and
dimensions of modules which, in contrast to the commutative case, are
no longer integer-valued. We order the submodules in a given partition
$(\bfp,\bfq)$ according to increasing dimension,
\beq
0<\dim {\cal E}_{p_1,q_1} \leq \dim {\cal E}_{p_2,q_2} \leq
\dim {\cal E}_{p_3,q_3}\leq \dots  \ .
\label{dimincrease}\eeq
Let $\nu_a$ be the number of submodules in this sequence that have the $a^{\rm
th}$ least dimension, which we denote by $\dim_a$. Then the integer
\beq
|\vec\nu\,|=\sum_{a\geq1}\nu_a
\label{nurational}\eeq
still gives the total number of submodules in a partition, and we may write the
partition function for rational $\theta$ as
\beq
Z_{p,q}(g^2,\theta)= \sum_{(\bfp,\bfq)\in{\cal P}_{p,q}(\theta)}
(-1)^{|\vec\nu\,|}\,\prod_{a\geq1}\frac{\Bigl(2g^2R^2\,(\dim_a)^3
\Bigr)^{-\nu_a/2}}{\nu_a!}~\e^{ - S(\bfp,\bfq;\theta)} \ .
\label{modYM}
\eeq
This expression provides a direct evaluation of Yang-Mills theory on a torus
with rational noncommutativity parameter $\theta$, without recourse to Morita
equivalence with the commutative theory. Note that in the case when all
submodules in the partitions have integer dimension, the formula (\ref{modYM})
reduces to that for the partition function of Yang-Mills theory on a
commutative torus in (\ref{commodYM}).

\subsection{Relation Between Commutative and Rational Noncommutative Gauge
Theories\label{ComNCYM}}

The arguments which led to the expression (\ref{modYM}) give an interesting way
to see the well-known connections between Yang-Mills theory on a noncommutative
torus with rational-valued $\theta$ and Yang-Mills theory defined on a
commutative torus. Consider the gauge theory defined on the Heisenberg module
${\cal E}_{p,q}$ over the noncommutative torus with deformation parameter
$\theta=n/s$, where $n$ and $s$ are relatively prime positive
integers. It can be verified from the definition  (\ref{dimcalEnm})
that any projective module over such a torus has a Murray-von~Neumann
dimension of at least $1/s$.
Since the total dimension of the module ${\cal E}_{p,q}$ is $p -nq/s$ in this
case, a partition $(\bfp,\bfq)$ which obeys the constraints (\ref{partrules})
and which consists of submodules of dimensions greater than or equal to $1/s$
has at most
\beq
\frac{p -nq/s}{1/s}= p s -   qn
\label{mostcomps}\eeq
components. Since Morita equivalence preserves the number of submodules in a
partition, any gauge theory which is dual to this rational noncommutative one
must admit partitions with $p s - qn$ components. On the other hand, for $U(N)$
Yang-Mills theory defined on a commutative torus, we know that due to the
constraints (\ref{partrules}), the maximum number of submodules in a partition
is $N$ (corresponding to $\nu_1=N$ and $\nu_a=0~~\forall a>1$). We conclude
that Yang-Mills theory on a Heisenberg module ${\cal E}_{p,q}$ over the
noncommutative torus with $\theta=n/s$ is Morita equivalent to a $U(N)$
commutative gauge theory of rank $N= ps -qn$. This result agrees with
how the rank of the noncommutative gauge theory appeared at the end of
section~\ref{Heisenberg}.

Notice that Morita equivalence maps submodules of the $U(ps-qn)$ commutative
gauge theory, as defined in the previous section, onto submodules of the
noncommutative gauge theory on the Heisenberg module ${\cal E}_{p,q}$ as
defined in section~\ref{Heisenberg}. The effect of this mapping on dimensions
of projective modules is to divide by $s$. This includes the irreducible
finite-dimensional representation of the Weyl-'t~Hooft algebra generated by the
$\hat Z_i$ in (\ref{NCcommrel}) as follows. The infinite-dimensional
center of the algebra
$\alg_{n/s}$ is generated by the elements $z_i=(\hat Z_i)^s$, $i=1,2$ which, in
an irreducible unitary representation, can be taken to be complex numbers of
unit modulus. The center can thereby be identified with the commutative algebra
$C^\infty({\bf T}^2)$ of smooth functions on the ordinary torus ${\bf T}^2$,
i.e. $\alg_{n/s}$ may be regarded as a twisted matrix bundle over
$C^\infty({\bf T}^2)$ of topological charge $n$ whose fibers are $s\times s$
complex matrix algebras $\mat_s$. In particular, there is a surjective algebra
homomorphism $\pi:\alg_{n/s}\to\mat_s$, sending the $\hat Z_i$ to the
corresponding $SU(s)$ shift and clock matrices, under which the entire center
of $\alg_{n/s}$ is mapped to $\complex$. In the language of Heisenberg modules
this representation corresponds to the finite-dimensional factor $\complex^s$
of the separable Hilbert space ${\cal E}_s=L^2({\bf T}^2)\otimes\complex^s$,
which allows for twisted boundary conditions on functions of the
ordinary torus ${\bf T}^2$ leading to the appropriate
Weyl-'t~Hooft algebra in this case. The irreducible finite-dimensional
representation of the algebra is thereby associated with a free module ${\cal
E}_s={\cal E}_{s,0}$ of vanishing Chern class. Therefore, the localization of
the partition function of quantum Yang-Mills theory on a rational
noncommutative torus is determined entirely by contributions from classical
solutions associated with Heisenberg modules as we have described them
above. By construction, this includes the Morita equivalent projective
modules over the ordinary torus.

\newsection{Yang-Mills Theory on a Noncommutative Torus: Irrational
Case\label{INCYM}}

Finally, we come to the case of irrational $\theta$. We claim that the formula
(\ref{modYM}) gives the Yang-Mills partition function as a sum over partitions
$(\bfp,\bfq)$ consisting of pairs of integers satisfying the constraints
(\ref{partrules}). Before justifying this claim, let us describe the
quantitative differences in the formula (\ref{modYM}) between the rational and
irrational cases. In fact, the analytical structure of the partition functions
in the two cases is very different due to the drastic differences of the
partitions in ${\cal P}_{p,q}(\theta)$ which contribute to the functional
integral. Recall from the previous section that in the case of rational
$\theta$, all modules have dimension at least $1/s$, and this fact was the crux
of the existence of the mapping between the rational and commutative gauge
theories. In contrast, when $\theta$ is irrational, submodules with arbitrarily
small dimension can contribute to a partition which characterizes a critical
point of the Yang-Mills action. As such, there is no {\it a priori} upper bound
on the number of submodules in a partition of ${\cal P}_{p,q}(\theta)$. While
for deformation parameter $\theta=n/s$ all partitions contain at most $ps-qr$
submodules of ${\cal E}_{p,q}$ of dimension at least $1/s$, in the irrational
case there are no such global bounds on the elements of ${\cal
P}_{p,q}(\theta)$. It is this fact that prevents Yang-Mills theory on a
noncommutative torus with irrational-valued $\theta$ from being Morita
equivalent to some commutative gauge theory of finite rank, and indeed
in this case the algebra $\alg_\theta$ has a trivial center.

As a consequence, in contrast to the rational case, the Yang-Mills partition
function on an irrational noncommutative torus receives contributions from
partitions containing arbitrarily many submodules.
However, it is possible to show that any partition
corresponding to a fixed finite action solution of the noncommutative
Yang-Mills equations of motion contains only finitely many components. By using
a Morita duality transformation (\ref{othertransfs}) we can transform the
action so that $\Phi=0$. Consider a partition $(\bfp,\bfq)\in{\cal
P}_{p,q}(\theta)$ on which the Yang-Mills action has the value
$S(\bfp,\bfq;\theta)=\xi\in\real_+$. Since (\ref{genymaction}) is a sum of
positive terms, this implies that $q_k^2\leq\xi\,(p_k-q_k\theta)$ for each
$k\geq1$. But the constraints (\ref{partrules}) imply
\beq
0<p_k-q_k\theta\leq p-q\theta
\label{pkineq}\eeq
for each $k\geq1$, and hence
\beq
q_k^2\leq\xi\,(p-q\theta) \ .
\label{qkineq}\eeq
{}From (\ref{qkineq}) it follows that $q_k$ can range over only a finite number
of integers, and hence from (\ref{pkineq}) the same is also true of $p_k$,
which establishes the result. In particular, we can pick out the minimum
dimension submodule ${\cal E}_{p_1,q_1}$ in a given partition
$(\bfp,\bfq)$ and order the submodules according to increasing
Murray-von~Neumann dimension as in (\ref{dimincrease}). The definition
(\ref{nurational}) still makes sense and hence so does the expression
(\ref{modYM}) for the partition function, provided that one now allows
for partitions with arbitrarily large (but finite) numbers of submodules.
Incidentally, this line of reasoning also shows that the set of values of the
noncommutative Yang-Mills action on the critical point set ${\cal
P}_{p,q}(\theta)$ is discrete, as is required of a Morse function
\cite{rieffel}.

Let us now indicate the reasons why (\ref{modYM}) is the correct result for the
partition function of Yang-Mills theory on a noncommutative torus with
irrational $\theta$. First of all, notice that the localization arguments of
section~\ref{partfnloc} which give the functional integral as a sum over
critical points of the Yang-Mills action are independent of the particular
value of
$\theta$. In a direct evaluation, the pre-exponential factors in (\ref{modYM})
would be determined by performing the functional Grassmann integrations and
taking the $t \rightarrow \infty$ limit in the localization formula
(\ref{ZNCYMlinlarget}). In this formula, $\theta$ is a continuous parameter and
we do not expect the calculation of contributions from Gaussian fluctuations to
depend on the rationality of $\theta$. Thus the pre-exponential factors in
(\ref{modYM}) yield the value of the fluctuation determinant at each critical
point in the semi-classical expansion of the partition function. Moreover, as
emphasized in section~\ref{ComNCYM}, the contributing submodules to this
expansion are always Heisenberg modules, which are the only projective
modules in the irrational case. In this regard it is interesting to note the
role of the alternating sign factors $(-1)^{|\vec\nu\,|}$ in
(\ref{modYM}). The global minimum of the action, which has
$|\vec\nu\,|=1$, is the only stable critical point of the
theory. According to general stationary phase arguments~\cite{locbook}, a
classical
solution with $n$ unstable modes is always weighted with a phase
$-\e^{\pi\ii n/2}$ in our normalization. Thus each submodule in a
partition which defines a critical point corresponds to a local
extremum of the noncommutative Yang-Mills action which is unstable in
two directions. Going back to the topological sum (\ref{partfnphys}),
we see that, as is the usual case in $U(p)$ gauge theory, each unit
charge instanton configuration yields $2p-2$ negative modes. The
instanton configurations will be studied in more detail in
section~\ref{moduli}.

Secondly, consider an approximation to the partition function for irrational
$\theta$ by rational theories. Formally, this requires a limit
$\theta=\lim_mn_m/s_m$ with both $n_m\to\infty$ and $s_m\to\infty$ as
$m\to\infty$. As we have seen in the previous section, the minimum dimension of
a submodule which is permitted over the noncommutative torus $\alg_{n_m/s_m}$
is $1/s_m$. Consequently, any rational approximation to the partition function
would contain partitions of arbitrarily small dimension, as we expect to see
for irrational values of the noncommutativity parameter $\theta$. With these
pieces of evidence at hand, we thereby propose that the partition function of
noncommutative gauge theory on a Heisenberg module ${\cal E}_{p,q}$ over a
two-dimensional torus is given for all values of the deformation parameter
$\theta$ by the expression
\bea
Z_{p,q}(g^2,\theta)&=&\sum_{(\bfp,\bfq)\in{\cal P}_{p,q}(\theta)}
(-1)^{|\vec\nu\,|}\,\prod_{a\geq1}\frac{\Bigl(2g^2R^2\,(p_a-q_a\theta)^3
\Bigr)^{-\nu_a/2}}{\nu_a!}\nn\\&&\times\,\prod_{k=1}^{|\vec\nu\,|}\exp\left[-
\frac1{2g^2R^2}\,
\Bigl(p_k - q_k\theta\Bigr)\left( \frac{ q_k}{p_k - q_k\theta}
- \frac{q}{p - q\theta} \right)^2\right] \ ,
\label{modYMtot}\eea
where the integer $a$ labels the $\nu_a$ submodules of dimension
$\dim_a=p_a-q_a\theta$. This formula exhibits the anticipated universality
between the irrational and rational cases~\cite{cr}, a feature which
we will see more of in the following. Note that the contributions from
classical solutions containing submodules of very small
Murray-von~Neumann dimension are
exponentially suppressed in (\ref{modYMtot}). In what follows we will explore
some applications of this formalism.

\newsection{Smoothness in $\theta$\label{smooth}}

An important issue surrounding noncommutative field theories in
general is the
behaviour of the partition function and observables as functions of the
noncommutativity parameter $\theta$. For example, the poles at
$\theta=0$ which
arise from perturbative expansions are the earmarks of the UV/IR mixing
phenomenon~\cite{UVIR}. However, it is not yet clear in the continuum
field theories
whether this is an artifact of perturbation theory or if it persists at a
nonperturbative level.\footnote{\baselineskip=12pt In the lattice
  regularization of noncommutative field theories~\cite{amns}, UV/IR mixing
  persists at a fully nonperturbative level as a kinematical effect.}
A clearer understanding of the behaviour of the nonperturbative theory
as a function of $\theta$ is therefore needed to fully address such
issues. Related to this problem is the question of approximation of
irrational noncommutative field theories by rational ones. If the
quantum field theory is at least continuous in $\theta$ then it can be
successively
approximated by rational theories. In particular, this would lead to a
hierarchy of Morita dual descriptions in terms of quasi-local degrees of
freedom~\cite{hash} and also finite-dimensional matrix model
approximations to the
continuum noncommutative field theory~\cite{amns1}. It has also been
suggested that smooth
behaviour of physical quantities in $\theta$ could, by Morita equivalence,
imply very stringent constraints on ordinary large $N$ gauge theories
on tori~\cite{gt}. These issues have been further addressed recently
in~\cite{barbon}.

While physically one would not expect to be able to measure a distinction
between rational-valued and irrational-valued observables, it has been observed
that in certain examples and at high energies, generic non-BPS
physical quantities exhibit discontinuous effects as functions of the
deformation parameter, due to the multifractal nature of the
renormalization group flows in these cases~\cite{epr}. For instance,
when $\theta$ is an irrational number, the cascade of Morita
equivalent descriptions is unbounded as the energy of the system
increases and no
quasi-local description of the theory is possible beyond a certain energy
level. To provide some different insight into these problems, in this section
we will analyze the behaviour of quantum Yang-Mills theory on the
noncommutative two-torus as a function of $\theta$, using its representation
(\ref{modYMtot}) as a sum over partitions associated with the Heisenberg module
${\cal E}_{p,q}$. As we have seen, each critical point of the Yang-Mills action
is determined by a partition which is a list of pairs of integers
$(\bfp,\bfq)=\Bigl\{ (p_k,q_k)\Bigr\}_{k=1}^{|\vec\nu\,|}$ labelling
submodules that obey the constraints (\ref{partrules}) on their
dimensions and Chern numbers. We will now develop a graphical
technique for constructing solutions of these constraints which will
serve as a useful method for obtaining solutions of the
noncommutative Yang-Mills equations of motion. This method makes no distinction
between rational or irrational $\theta$ and smoothly interpolates between the
two cases. We will then use it to prove the {\it smoothness} of the Yang-Mills
partition function (\ref{modYMtot}) as a function of the noncommutativity
parameter $\theta$. This continuity result is in agreement with an analysis,
based on continued fraction approximations, of the behaviour of classical
averages on a fixed projective module~\cite{lls}.

\subsection{Graphical Determination of Classical Solutions}

Consider the integral lattice ${\rm K}_0(\alg_\theta)$ of K-theory
charges, which we will view as a subset of the plane $\real^2$. Each
point $(p_k,q_k)$ on this lattice corresponds to an isomorphism class
${\cal E}_{p_k,q_k}$ of projective modules over the noncommutative
torus. Through each such point we draw a line
in $\real^2$ of constant (positive) dimension according to the equation
\beq
p-  q\theta = p_k -q_k\theta  \ , ~~ k=1,2,3,\dots \ .
\label{constdimline}\eeq
These lines all have slope $\theta^{-1}$. For irrational values of $\theta$,
there is a unique solution, $(p,q)=(p_k,q_k)$, to (\ref{constdimline}) for each
$k$, and hence there is only one point of the integer lattice on each line.
Consequently there are an infinite number of parallel lines of
constant dimension in any region of the K-theory lattice. On the other hand, if
$\theta=n/s$ is a rational number, then there are infinitely many solutions
$(p,q)$ of the equation (\ref{constdimline}) for each $k$ and hence a large
degeneracy of lattice points lying on each line. In this case there are only
a finite number of lines of constant dimension in any region of the K-theory
lattice.

\begin{figure}[htb]
\epsfxsize=3in
\bigskip
\centerline{\epsffile{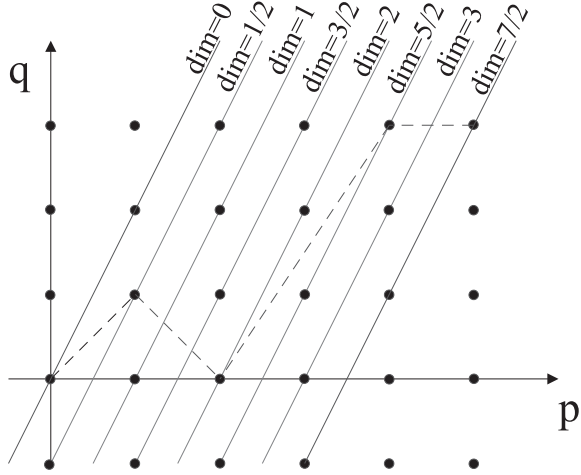}}
\caption{\baselineskip=12pt{\it Graphical representation of the partition
$\Bigl\{(1,1)\,,\,(1, -1)\,,\, (2,3)\,,\, (1,0)\Bigr\}\in{\cal
  P}_{5,3}(\theta)$ which defines a solution of the noncommutative
Yang-Mills equations of motion on the projective module ${\cal
  E}_{5,3}$. The sequence of lines in $\real^2$ of constant dimension
for the case $\theta =1/2$ is depicted. The dashed line goes through
the sequence of points whose successive differences make up the
  elements of the partition.}}
\bigskip
\label{smoothfig}\end{figure}

For a given Heisenberg module ${\cal E}_{n,m}$, there are two important lines
of constant dimension which will enable the enforcing of the constraints
(\ref{partrules}).  These are the lines $ p- q\theta =0$ and $ p-  q\theta= n-
m\theta$. A partition which yields a critical point of the Yang-Mills action on
${\cal E}_{n,m}$ is found by taking a sequence of points lying on lines of
strictly increasing dimension, beginning at the origin $(0,0)$ and terminating
at the point $(n,m)$. Taking the difference of the coordinates of successive
points gives the topological numbers $(n_k,m_k)$ of the submodules in the
partition.
The choice of a sequence of points which lie on lines of strictly
increasing dimension guarantees that each submodule is of positive dimension.
Fixing the initial and final points ensures that the constraints on the total
dimension and Chern number are satisfied.
An illustrative example of this procedure for the module
${\cal E}_{5,3}$ is depicted in  Figure~\ref{smoothfig}.
All finite sequences of points obeying these rules give
all possible solutions of the constraints (\ref{partrules}), and hence all
critical points of the noncommutative Yang-Mills action corresponding to all
solutions of the equations of motion. Note that the integer
$|\vec\nu\,|$, counting the total number of submodules in a partition,
may in this way be regarded as a topological invariant of the
associated graphs.

\subsection{Proof of $\theta$-Smoothness}

Having determined all partitions graphically for fixed $\theta$, we can now
study how semi-classical quantities vary with a change of $\theta$. From $n
-m\theta  =\dim {\cal E}_{n,m}$, we see that $\theta$ is the inverse slope of
lines of constant dimension in the $(p,q)$ plane. Thus a change in $\theta$
amounts to a change in slope of the lines of constant dimension. For
partitions a small change in $\theta$ leads to a small change in the
dimensions of submodules in a partition but leaves the number
$|\vec\nu\,|$ of submodules and their topological numbers
unchanged. The partition function (\ref{modYMtot}) clearly varies
smoothly under such variations of the noncommutativity parameter. This
smooth behaviour terminates when a change in $\theta$ leads to a
violation of the requirement that each submodule of a partition be of
positive dimension. Such a condition can occur when a submodule of
very small dimension to
the right of the line $p-q\theta=0$ is pushed through to negative
dimension by an infinitesimal variation of $\theta$. For example, the
partition depicted in
Figure~\ref{smoothfig} represents a valid solution  of Yang-Mills theory
on the module ${\cal E}_{5,3}$ for all $\theta<1$.  As $\theta$
approaches unity,  the dimension  of the first submodule ${\cal E}_{1,1}$
vanishes. Thus at $\theta=1$, the constraints (\ref{partrules})
defining partitions are violated and this partition is abruptly removed
from the list ${\cal P}_{5,3}(\theta)$ of partitions
which contribute to the partition function. Of course such an elimination
occurs for any partition containing a submodule of vanishing
dimension\footnote{\baselineskip=12pt Recall that the components of a
  partition are partially ordered according to increasing submodule dimension.}
and it would appear in general that this leads to a
discontinuity in the partition function as a function of
$\theta$. There are also various ``degenerate'' cases that appear to
lead to discontinuities, such as those partitions for which the
dimension of a component which doesn't appear first in the list
vanishes, or those where the dimensions of two submodules become equal
when $\theta$ is varied. This leads to a reordering of the submodules
and therefore a discontinuous change in the graphical representation
of the previous subsection. However, these latter cases do not affect
the partition sum in a discontinuous way, and thus only the former
types of discontinuities appear to remain.

In fact this is not the case and the partition function is smooth in $\theta$.
The reason is that the contribution to the partition function (\ref{modYMtot})
from partitions with submodules
of vanishing dimension are exponentially suppressed, since the
Boltzmann weight associated with such topological numbers $(n,m)$ is of
order $\e^{-1/\dim{\cal E}_{n,m}}$. Consequently, the partition function
has already exponentially damped any contribution from a partition before it is
discontinuously dropped due to the positive dimension constraint. It is easy to
see that even though derivatives of the partition function with respect to
$\theta$ will generate singular pre-exponential factors when submodule
dimensions vanish, these singularities are all trumped by exponential
suppression from the action. Thus all derivatives of the partition function
with respect to $\theta$ are also finite and continuous. Note that, in the
context of rational approximations to irrational values of the noncommutativity
parameter, this analysis also shows that perturbations about any rational value
of $\theta$ will miss exponentially small contributions to the partition
function, which may be related to some of the peculiarities observed in the
rational approximations of irrational noncommutative gauge theories~\cite{epr}.

The $\theta$-smoothness proof can also be extended to physical (gauge
invariant) observables which are at most polynomially singular in
$\theta$ for modules of vanishing dimension. One such class of
observables are the ``topological'' observables obtained by
differentiating the partition function (\ref{ZNCYMlin}) with respect
to the Yang-Mills coupling constant,
\bea
\left(\frac\partial{\partial g^2}\right)^n\ln Z_{p,q}(g^2,\theta)&=&
\left(\frac1{8\pi^2R^2}\right)^n\,\left\langle\left(\hat\phi\,,\,\hat
\phi\right)^n\right\rangle_{\rm conn}\nn\\&=&\left(\frac1{8\pi^2R^2}\right)^n\,
\left\langle\,\prod_{r=1}^n\tr^{~}_N\,\phi(x_r)^2\right\rangle_{\rm conn} \ ,
\label{topobs}\eea
where the brackets $\langle\cdots\rangle^{~}_{\rm conn}$ denote connected
correlation functions with respect to the functional integral
(\ref{ZNCYMlin}), and $x_1,\dots,x_n$ are arbitrary points on ${\bf
  T}^2$. For $n=1$ this observable is proportional to the average
energy of the system $\left\langle\Tr^{~}_{{\cal E}_{p,q}}\,\hat
  F_{\hat A}^2\right\rangle$ on the Heisenberg module~${\cal E}_{p,q}$.

\newsection{Instanton Moduli Spaces\label{moduli}}

The expansion (\ref{modYMtot}) of the partition function of gauge theory on a
noncommutative torus has a natural interpretation as a sum over noncommutative
instantons in two dimensions, in the sense that we have defined them at the
beginning of section~\ref{InstClass}. They are classified topologically by the
homotopy classes $(p,q)$ of the space of Hermitian projectors
$\hil^{~}_\theta$,
and they are inherently nonperturbative since their action contribution to the
path integral is of order $\e^{-1/g^2}$. However, the semi-classical expansion
does not organize the contributions from classical solutions into gauge orbits.
Different partitions $(\bfp,\bfq)$ may give contributions which
should be identified as coming from the same instanton. In
this section we will discuss the rearrangement of the series (\ref{modYMtot})
into a sum over gauge inequivalent critical points and describe the structure
of the moduli spaces of instantons that arise, comparing them with those of
ordinary Yang-Mills theory in two dimensions.

\subsection{Weak Coupling Limit\label{Weak}}

We will begin with the weak coupling limit $g^2\to0$ of noncommutative gauge
theory as it is the simplest case. Due to the invariance property
(\ref{omegainv}), the moduli space ${\cal M}_{p,q}(\theta)$ of constant
curvature connections on the Heisenberg module ${\cal E}_{p,q}$ modulo
noncommutative gauge transformations has a natural symplectic structure
inherited from the symplectic two-form (\ref{omegadef}) on $\CC_{p,q}$. As
shown in section~\ref{locNCYM}, the partition function $Z_{p,q}(g^2=0,\theta)$
formally computes the symplectic volume of ${\cal M}_{p,q}(\theta)$. Let us
first describe this moduli space~\cite{cr}. By using a Morita duality
transformation
(\ref{othertransfs}) of the background magnetic flux $\Phi$ if necessary, we
can assume that $f\neq0$ in (\ref{constcurvcond}) without loss of generality.
We therefore need to classify the irreducible representations determined by the
Heisenberg module (\ref{calEnmexpl}). As discussed in section~\ref{Heisenberg},
the Weyl-'t~Hooft algebra (\ref{WeyltHooftalg}) has $N$ irreducible components,
where $N$ is the rank of the noncommutative gauge theory given by
(\ref{Ngcdnm}). On the other hand, in section~\ref{ComNCYM} we saw that each
such irreducible representation has a pair of complex moduli $(z_1,z_2)$
generating the center of the Weyl-'t~Hooft algebra. Thus the inequivalent
irreducible representations of the matrix algebra (\ref{WeyltHooftalg}) are
labelled by a pair of complex numbers $\zeta=(z_1,z_2)\in\tilde{\bf T}^2$ which
live on a commutative torus dual to the original noncommutative torus.

If ${\cal W}_\zeta\subset\complex^q$, $\zeta\in\tilde{\bf T}^2$ are the
corresponding irreducible representations, then the Heisenberg module
(\ref{calEnmexpl}) decomposes into irreducible $\alg_\theta$-modules according
to
\beq
{\cal E}_{p,q}=L^2(\real)\otimes\left({\cal W}_{\zeta_1}\oplus\cdots\oplus
{\cal W}_{\zeta_N}\right) \ .
\label{calEpqirrep}\eeq
Gauge transformations which preserve the constant curvature condition
(\ref{constcurvcond}) are finite-dimensional unitary matrices in $U(q)$.
Central elements of the Weyl-'t~Hooft algebra are represented by diagonal
matrices with respect to the decomposition (\ref{calEpqirrep}). There is a
residual gauge symmetry which acts by permutation of the $N$ summands in
(\ref{calEpqirrep}) as the permutation group $\Sigma_N$, and therefore the
moduli space of constant curvature connections associated with the module
${\cal E}_{p,q}$ over the noncommutative torus is the symmetric
orbifold~\cite{cr}
\beq
{\cal M}_{p,q}(\theta)={\rm Sym}^N\,\tilde{\bf T}^2\equiv
\left(\tilde{\bf T}^2\right)^N\,/\,\Sigma_N
\label{calMpq}\eeq
of dimension $2N$. This space is identical to the moduli space of flat bundles
for commutative Yang-Mills theory on an elliptic Riemann surface with structure
group $U(N)$~\cite{AB}, i.e. ${\cal M}_N(\theta=0)={\rm
Hom}\Bigl(\pi^{~}_1({\bf T}^2)\,,\,U(N)\Bigr)\,/\,U(N)$, since the maximal
torus of $U(N)$ is $U(1)^N$
consisting of diagonal matrices and its discrete Weyl subgroup is precisely the
symmetric group $\Sigma_N$. The standard symplectic geometry on (\ref{calMpq})
possesses conical singularities on the coincidence locus, i.e. the ``diagonal''
subspace of $\left(\tilde{\bf T}^2\right)^N$.

Let us now consider this result in light of the instanton expansion. We
take $\Phi=0$ without loss of generality. The $g^2\to0$ limit of the Boltzmann
factor in the partition function (\ref{modYMtot}) is non-vanishing only in the
zero instanton sector $q_k=0~~\forall k\geq1$. But the constraints
(\ref{partrules}) with all $q_k=0$ are just equivalent to those we encountered
in section~\ref{CYM} in the commutative limit $\theta=0$, with rank $N=p$. The
same is true of the fluctuation determinant factors in (\ref{modYMtot}), and
hence the partition function at weak coupling is given by
\bea
Z_{p,q}(g^2=0,\theta)~=~\lim_{g^2\to0}~\sum_{\vec\nu\,\,:\,\sum_aa\nu_a=N}~
\prod_{a=1}^N\frac{(-1)^{\nu_a}}{\nu_a!}\,\Bigl(2g^2R^2a^3
\Bigr)^{-\nu_a/2} \ ,
\label{partweak}\eeq
where we have eliminated the (constant) exponential factor by a suitable
renormalization of the quantum field theory~\cite{Wittenloc}. The independence
of
(\ref{partweak}) in the noncommutativity was observed in section~\ref{locNCYM},
where we saw that the auxiliary field $\phi$ in (\ref{ZNCYMlin}) was
essentially a commutative field. In this way, the theory at $g^2=0$ eliminates
all dependence on the parameters $\theta$ and $R$, and it is identical to
topological Yang-Mills theory on the commutative two-torus~\cite{Wittenloc}.
This feature of the weakly coupled gauge theory is in agreement with the
coincidence of the moduli space of the zero instanton sector in the commutative
and noncommutative cases. It should be stressed though that, in contrast to the
commutative case, by Morita equivalence the expressions for the moduli space
(\ref{calMpq}) and partition function (\ref{partweak}) over the noncommutative
torus hold generically for {\it all} (not necessarily flat) constant curvature
connections. In other words, in the noncommutative case the gauge quotiented
level sets of the moment map $\mu$ on $\CC_{p,q}$ are all equivalent.

The partition function has non-analytic behaviour in the Yang-Mills coupling
constant as $g^2\to0$, with a pole of order $|\vec\nu\,|$ in $g$ for each
partition. We can relate the singularities arising in (\ref{partweak}) in a
very precise way to the orbifold singularities of the moduli space
(\ref{calMpq}) which appear whenever the Heisenberg module ${\cal E}_{p,q}$ is
reducible. The latter singular points come from the fixed point set of the
action of the permutation group $\Sigma_N$ on $\left(\tilde{\bf T}^2\right)^N$,
which is straightforward to describe. As in section~\ref{CYM}, let
$C[\vec\nu\,]=C[\nu_1,\dots,\nu_N]$ be the conjugacy class of a given element
$\sigma\in\Sigma_N$. An elementary cycle of length $a$ leaves an $N$-tuple
$(\zeta_1,\dots,\zeta_N)\in\left(\tilde{\bf T}^2\right)^N$ invariant only if
the $a$ points on which it acts coincide. It follows that the fixed point locus
of any permutation $\sigma$ in the conjugacy class $C[\vec\nu\,]$ is given by
\beq
\left[\left(\tilde{\bf T}^2\right)^N\right]^{C[\vec\nu\,]}~=~
\prod_{a=1}^N\left(\tilde{\bf T}^2\right)^{\nu_a} \ .
\label{fixedpointlocus}\eeq
On each such fixed point set there is still the action of the stabilizer
subgroup ${\cal C}(\vec\nu\,)$ of $\Sigma_N$, which consists of all
elements $\sigma'\in\Sigma_N$ that commute with $\sigma$ and is given
explicitly in terms of semi-direct products as
\beq
{\cal C}(\vec\nu\,)=\prod_{a=1}^N\Sigma_{\nu_a}
\ltimes\zed_a^{\nu_a} \ .
\label{calCsigma}\eeq
Here the symmetric group $\Sigma_{\nu_a}$ permutes the $\nu_a$ cycles of length
$a$, while each cyclic group $\zed_a$ acts within one particular cycle of
length $a$. Distinct singular points of the symmetric orbifold (\ref{calMpq})
then arise at the ${\cal C}(\vec\nu\,)$ invariants of the fixed point
loci (\ref{fixedpointlocus}). Only the subgroups $\Sigma_{\nu_a}$ of the
centralizer (\ref{calCsigma}) act non-trivially on (\ref{fixedpointlocus}). The
singular point locus of the moduli space of constant curvature connections is
thereby obtained as the disjoint union over the conjugacy classes
$C[\vec\nu\,]$ of $\Sigma_N$ of the strata
\beq
\left[\left(\tilde{\bf T}^2\right)^N\right]^{C[\vec\nu\,]}~/~
{\cal C}(\vec\nu\,)~=~\prod_{a=1}^N\,{\rm Sym}^{\nu_a}\,\tilde{\bf T}^2 \ .
\label{singlocus}\eeq

Given the result (\ref{singlocus}), the interpretation of the singularities in
the formal orbifold volume (\ref{partweak}) is now clear. It is a sum over the
connected components, labelled by conjugacy classes in $\Sigma_N$, of the total
singular locus of the orbifold symmetric product (\ref{calMpq}). Within each
conjugacy class (partition) $C[\vec\nu\,]$, the contribution from a gauge
equivalence class of connections associated with a toroidal factor
$\left(\tilde{\bf T}^2\right)^{\nu_a}$ in (\ref{singlocus}) is weighted by a
singular fluctuation determinant $(-1)^{\nu_a}\,(2g^2R^2a^3)^{-\nu_a/2}$.
Recall from section~\ref{RatPart} that the module dimension factor $a^3$ here
ensures invariance under Morita duality. Gauge invariance dictates that the
total
contribution from the $\nu_a$ cycles of length $a$ (submodules of rank $a$) be
divided by the appropriate residual symmetry factor $\nu_a!$ which is the order
of the local orbifold group $\Sigma_{\nu_a}$ acting in (\ref{singlocus}). Thus
the conical singularities of the zero instanton sector are not smoothed out by
the noncommutativity, as one might have naively
expected~\cite{schwarzinst,headrick}, and the moduli spaces of flat
connections are the same in both commutative and noncommutative cases. The
corresponding partition functions (\ref{partweak}) represent the contribution
of the global minimum $\mu^{-1}(0)$ to the localization formula for the
functional integral. We shall now analyze how these properties change as one
moves away from the weak coupling limit of the noncommutative gauge theory. As
we will see, the orbifold singularities for coincident instantons on the moduli
space still persist. Geometrically, the noncommutative instantons of two
dimensional gauge theory on a torus remain point-like and hence have no
smoothing effect on the conical singularities that occur on ${\rm
Sym}^N\,\tilde{\bf T}^2$ where two or more points come together.

\subsection{Instanton Partitions\label{InstConf}}

To count instantons labelled by a generic partition $(\bfp,\bfq)$ consisting of
non-zero Chern numbers $q_k$, we need to arrange the expansion (\ref{modYMtot})
into a sum over gauge inequivalent classical solutions. The essential problem
which arises is the isomorphism ${\cal E}_{mp,mq}\cong\oplus^m{\cal E}_{p,q}$
of
projective modules. Partitions of either side of this isomorphism lead to gauge
equivalent contributions to the partition function and, in particular, from
(\ref{constcurvcondF}) it follows that the minimizing connections on ${\cal
E}_{mp,mq}$ and ${\cal E}_{p,q}$ have the same constant curvature. Thus we need
to refine the definition of partition given in section~\ref{Critical} somewhat
so as to combine submodules which yield the same constant curvature and hence
prevent the over-counting of distinct noncommutative Yang-Mills stationary
points~\cite{rieffel}. This we do by writing any submodule dimension in the
form
\beq
p_k-q_k\theta=N_k\left(p_k'-q_k'\theta\right)
\label{Nkprimes}\eeq
where $N_k={\rm gcd}(p_k,q_k)$, and the integers $p_k'$ and $q_k'$ are
relatively prime. The corresponding curvature (\ref{constcurvcondF}) is
independent of the noncommutative rank $N_k$, and so we should also restrict to
submodules for which each K-theory charge $(p_k',q_k')$ is distinct. Therefore,
we restrict the counting of critical points of the noncommutative Yang-Mills
action to the sets of integers
$(\bfN,\bfp',\bfq')\equiv\Bigl\{(N_a,p_a',q_a')\Bigr\}_{a\geq1}$ which satisfy,
in addition to the constraints (\ref{partrules}), the requirements that
$N_a>0$, $p_a'$ and $q_a'$ are relatively prime, and the pairs of integers
$(p_a',q_a')$ are all distinct. We shall refer to such a collection of integers
as an ``instanton partition''. The additional constraints imposed on an
instanton partition guarantee that we do not count as distinct those partitions
which contain some submodules that can themselves be decomposed into
irreducible components.

Let us look at the structure of the moduli space ${\cal
M}_{p,q}(\bfN,\bfp',\bfq';\theta)$ associated with an instanton partition
$(\bfN,\bfp',\bfq')$~\cite{rieffel}. We want to
determine the space of gauge orbits of the associated critical point
connections $\hat\nabla^{\rm cl}=\bigoplus_{a\geq1}\hat\nabla_{(a)}^{\rm c}$ on
submodule decompositions
\beq
{\cal E}_{p,q}=\bigoplus_{a\geq1}{\cal E}_{N_ap_a',N_aq_a'} \ .
\label{calEpqinstpart}\eeq
Since each constant curvature on ${\cal
E}_{N_ap_a',N_aq_a'}$ is distinct and any gauge transformation $\hat U\in{\sf
G}({\cal E}_{p,q})$ preserves the constant curvature conditions, every $\hat U$
is also a unitary operator on each instanton submodule ${\cal
E}_{N_ap_a',N_aq_a'}\to{\cal E}_{N_ap_a',N_aq_a'}$. It follows that the
instanton moduli space is given by
\beq
{\cal M}_{p,q}(\bfN,\bfp',\bfq';\theta)=\prod_{a\geq1}{\cal
M}_{N_ap_a',N_aq_a'}(\theta) \ ,
\label{calMpqprod}\eeq
where each ${\cal M}_{N_ap_a',N_aq_a'}(\theta)$ is the moduli space of constant
curvature connections on the Heisenberg module ${\cal E}_{N_ap_a',N_aq_a'}$.
{}From (\ref{calMpq}) we thus find that (\ref{calMpqprod}) can be written in
terms of a product of symmetric orbifolds as~\cite{rieffel}
\beq
{\cal M}_{p,q}(\bfN,\bfp',\bfq';\theta)=\prod_{a\geq1}\,{\rm Sym}^{N_a}\,
\tilde{\bf T}^2 \ .
\label{instmodspace}\eeq
This result generalizes the instanton moduli space (\ref{calMpq}) which
corresponds to the global minimum of the noncommutative Yang-Mills action on
${\cal E}_{p,q}$.

\subsection{Examples\label{Charge1}}

To get a feel for how the moduli spaces (\ref{instmodspace}) classify the
reorganization of the partition function (\ref{modYMtot}) into a sum over
distinct instanton contributions, let us consider two very simple
examples~\cite{griguolo1}. For $\theta=0$ and $p=2$ the partition function is
easily written in the form
\beq
Z_{2,q}(g^2,0)=-\frac{\e^{-q^2/4g^2R^2}}{\sqrt{16g^2R^2}}+\frac1{4g^2R^2}\,
\sum_{q_1=-\infty}^\infty\e^{-\frac1{2g^2R^2}\,\bigl(q_1^2+(q-q_1)^2\bigr)} \ .
\label{Z2qodd}\eeq
When the Chern number $q$ is odd, this is a sum over inequivalent instanton
configurations, and the two terms in (\ref{Z2qodd}) are associated respectively
with the smooth moduli spaces
\bea
{\cal M}_{2,q}(1,2,q;0)&=&\tilde{\bf T}^2 \ , \label{calM2qodd}\\{\cal M}_{2,q}
\Bigl((1,1,q_1)\,,\,(1,1,q_2)\,;\,0
\Bigr)&=&\tilde{\bf T}^2\times\tilde{\bf T}^2 \ .
\label{calMqodd}\eea
Heuristically, with the appropriate symmetry factors, each factor $\tilde{\bf
T}^2$, representing the single instanton moduli space, contributes a mode with
fluctuation determinant $-1/\sqrt{16g^2R^2}$. On the other hand, when $q=2q'$
is even there is a term in the infinite series in (\ref{Z2qodd}) which yields
the same Boltzmann weight as the first term, and so these two terms should be
combined to give
\beq
Z_{2,2q'}(g^2,0)=\e^{-q'^2/g^2R^2}\,\left(-\frac1{\sqrt{16g^2R^2}}+\frac1
{4g^2R^2}\right)+\frac1{4g^2R^2}\,\sum_{q_1\neq q'}\e^{-\frac1{2g^2R^2}\,
\bigl(q_1^2+(2q'-q_1)^2\bigr)} \ . \nn\\
\label{Z2qeven}\eeq
Again the last term in (\ref{Z2qeven}) may be attributed to contributions from
instantons in the smooth moduli space (\ref{calMqodd}) with $q=2q'$ and
$q_1\neq q_2$. The two gauge equivalent instanton contributions to the first
term are attributed with the singular moduli space in this case,
\beq
{\cal M}_{2,2q'}(2,1,q';0)={\rm Sym}^2\,\tilde{\bf T}^2 \ .
\label{calMqextra}\eeq
The singular locus of the symmetric orbifold (\ref{calMqextra}) is ${\rm
Sym}^2\,\tilde{\bf T}^2\,\amalg\,\tilde{\bf T}^2$ with the disjoint sets
corresponding to the identity and order two elements of the cyclic group
$\zed_2$, respectively. As in section~\ref{Weak}, the sum of contributions to
the first term in (\ref{Z2qeven}) are readily seen to be those associated with
the components of the total singular point locus of (\ref{calMqextra}).

For $\theta=0$ and $p=3$ the partition function is given by
\bea
Z_{3,q}(g^2,0)&=&-\frac{\e^{-q^2/6g^2R^2}}{\sqrt{54g^2R^2}}+\frac1{32g^2R^2}\,
\sum_{q_1=-\infty}^\infty\e^{-\frac1{4g^2R^2}\,\bigl(2q_1^2+(q-q_1)^2\bigr)}
\nn\\&&-\,\frac1{6(2g^2R^2)^{3/2}}\,\sum_{q_1=-\infty}^\infty~
\sum_{q_2=-\infty}^\infty\e^{-\frac1{2g^2R^2}\,\bigl(q_1^2+q_2^2+(q-q_1-q_2)^2
\bigr)} \ .
\label{Z3qgen}\eea
For any $q\notin3\,\zed$ the expression (\ref{Z3qgen}) can be written as a sum
over distinct instanton contributions as
\bea
Z_{3,q}(g^2,0)&=&-\frac{\e^{-q^2/6g^2R^2}}{\sqrt{54g^2R^2}}+\left(
\frac1{32g^2R^2}-\frac1{6(2g^2R^2)^{3/2}}\right)\,
\sum_{q_1=-\infty}^\infty\e^{-\frac1{4g^2R^2}\,\bigl(2q_1^2+(q-q_1)^2\bigr)}
\nn\\&&-\,\frac1{6(2g^2R^2)^{3/2}}\,\sum_{q_1=-\infty}^\infty~
\sum_{q_2\neq q_1}\e^{-\frac1{2g^2R^2}\,\bigl((2q_1-q)^2+(2q_1-q_2)^2+q_2^2
\bigr)}\nn\\&&-\,\frac1{6(2g^2R^2)^{3/2}}\,\sum_{q_1\neq q\,{\rm mod}\,2}~
\sum_{q_2=-\infty}^\infty\e^{-\frac1{2g^2R^2}\,\bigl(q_1^2+q_2^2+(q-q_1-q_2)^2
\bigr)}
\label{Z3qdistinct}\eea
corresponding respectively to the instanton moduli spaces
\bea
{\cal M}_{3,q}(1,3,q;0)&=&\tilde{\bf T}^2 \ , \label{calM3q1}\\
{\cal M}_{3,q}\Bigl((1,1,q_1)\,,\,(2,1,q_2)\,;\,0\Bigr)&=&\tilde{\bf T}^2
\times{\rm Sym}^2\,\tilde{\bf T}^2 \ , \label{calM3q2}\\
{\cal M}_{3,q}\Bigl((1,1,q_1)\,,\,(1,2,q_2)\,;\,0\Bigr)&=&\tilde{\bf T}^2
\times\tilde{\bf T}^2 \ , \label{calM3q2odd}\\
{\cal M}_{3,q}\Bigl((1,1,q_1)\,,\,(1,1,q_2)\,,\,(1,1,q_3)\,;\,0\Bigr)&=&
\tilde{\bf T}^2\times\tilde{\bf T}^2\times\tilde{\bf T}^2 \ .
\label{calM3q3}\eea
Note again how the fluctuation determinants in (\ref{Z3qdistinct}) weight each
factor of $\tilde{\bf T}^2$ in the corresponding moduli space, and how the
second term incorporates the sum over singularities of the symmetric orbifold
${\rm Sym}^2\,\tilde{\bf T}^2$ in (\ref{calM3q2}). For $q=3q'$, the second term
in (\ref{Z3qdistinct}) yields a contribution to the global minimum for
$q_1=q'$, and we have
\bea
Z_{3,3q'}(g^2,0)&=&\e^{-3q'^2/2g^2R^2}\,\left(-\frac1{\sqrt{54g^2R^2}}+
\frac1{32g^2R^2}-\frac1{6(2g^2R^2)^{3/2}}\right)\nn\\&&+\,\left(
\frac1{32g^2R^2}-\frac1{6(2g^2R^2)^{3/2}}\right)\,
\sum_{q_1\neq q'}\e^{-\frac1{4g^2R^2}\,\bigl(2q_1^2+(3q'-q_1)^2\bigr)}
\nn\\&&-\,\frac1{6(2g^2R^2)^{3/2}}\,\sum_{q_1=-\infty}^\infty~
\sum_{q_2\neq q_1}\e^{-\frac1{2g^2R^2}\,\bigl((2q_1-3q')^2+(2q_1-q_2)^2+q_2^2
\bigr)}\nn\\&&-\,\frac1{6(2g^2R^2)^{3/2}}\,\sum_{q_1\neq3q'\,{\rm mod}\,2}~
\sum_{q_2=-\infty}^\infty\e^{-\frac1{2g^2R^2}\,
\bigl(q_1^2+q_2^2+(3q'-q_1-q_2)^2\bigr)} \ .
\label{Z33qprime}\eea
The last three terms in (\ref{Z33qprime}) may again be attributed to
contributions associated with the instanton moduli spaces
(\ref{calM3q2})--(\ref{calM3q3}), respectively, with $q=3q'$ and $q_1\neq
q_2\neq q_3$. The first term represents the gauge equivalent instanton
contributions coming from replacing the smooth moduli space (\ref{calM3q1}) by
the singular one
\beq
{\cal M}_{3,3q'}(3,1,q';0)={\rm Sym}^3\,\tilde{\bf T}^2 \ ,
\label{calM33qprime}\eeq
with each fluctuation determinant associated with the singular points of the
orbifold (\ref{calM33qprime}) corresponding to the three conjugacy classes of
the symmetric group $\Sigma_3$.

These two simple examples illustrate the general technique involved in
reorganizing the sum (\ref{modYMtot}) over critical points into distinct
instanton contributions. They can be deduced, as above, from the singularity
structures of the totality of instanton moduli spaces (\ref{instmodspace})
corresponding to a Heisenberg module. The Boltzmann weight associated with an
instanton partition $(\bfN,\bfp',\bfq')$ is given by
\beq
\e^{-S(\bfN,\bfp',\bfq';\theta)}=\prod_{a\geq1}\e^{-\frac1{2g^2R^2}\,
N_aq_a'^2/(p_a'-q_a'\theta)} \ ,
\label{Boltzmanninst}\eeq
and about it there are a finite number of quantum fluctuations
representing a finite, but non-trivial, perturbative expansion in
$g^{-1}$. These fluctuations are determined by the singular locus of
the corresponding symmetric orbifolds in (\ref{instmodspace}). The
combinatorial problem of summing over all such instanton partitions
is in general quite involved, especially for irrational values of
$\theta$ when there are infinitely many partitions. However, repeating
analogous arguments to those around (\ref{pkineq},\ref{qkineq}) shows
that an instanton partition contains only finitely many
components. Thus the perturbative expansion around each instanton
contribution contains only finitely many terms, although in the
irrational case the exponential prefactor is no longer a polynomial of
set order. It is amusing that, within the class of noncommutative
gauge theories, Morita equivalence allows such a moduli space
classification of the instanton contributions even in the commutative
case. Such a characterization is otherwise not possible because one
only knows the structure of the moduli space of {\it flat} connections
of commutative gauge theory on ${\bf T}^2$. Notice also that for
$\theta\neq0$ the instanton sums are no longer given by elementary
theta-functions.

Finally, let us note that the instantons which contribute to the
semi-classical expansion of noncommutative gauge theory that we have
developed are reminiscent of the solitons on noncommutative tori which
arise as solutions of open string field theory describing unstable
D-branes wrapping a two-dimensional torus in the background of a constant
$B$-field~\cite{moore,kraj}. An extremum of the tachyon potential is described
by a projector of the algebra $\alg_\theta$, and leads to an effective
gauge theory on the corresponding projective module determined by the
tachyon. The remaining string field equations of motion are then
solved by direct sum decompositions of the given Heisenberg module as
we have described them in this paper. A special instance of this are
the fluxon solutions which describe the finite energy instantons,
carrying quantized magnetic flux, of gauge theory on the noncommutative
plane~\cite{poly}--\cite{blp}. In the present setting these are the classical
solutions associated with partitions consisting of only the full
module, giving the global minimum of the Yang-Mills action. For the
module ${\cal E}_{p,q}$, these solutions have gauge field strength
(\ref{constcurvcondF}) and partition function
\beq
Z_{N,q}^{\rm fluxon}(g^2,\tilde\theta)=\frac{\e^{-Nq'^2/2g^2R^2(p'-q'\theta)}}
{\sqrt{2g^2R^2N^3(p'-q'\theta)^3}} \ .
\label{Zfluxon}\eeq
In the large area limit $R\to\infty$ with the dimensionful
noncommutativity parameter $\tilde\theta=2\pi R^2\,\theta$ finite,
this is the contribution to the functional integral, along with the
appropriate Gaussian fluctuation factor, from a fluxon of magnetic
charge $q=Nq'$ in gauge group rank $N$. The sum over all $q\in\zed$ in this
limit determines the expansion of noncommutative gauge theory on
$\real^2$ in terms of fluxons~\cite{gsv}.

\subsection*{Acknowledgments}

We thank C.-S.~Chu, J.~Gracia-Bond\'{\i}a, G.~Landi, F.~Lizzi, B.~Schroers and
G.~Semenoff for helpful discussions and correspondence. We also thank
the referee for prompting us to correct several technical
inconsistencies in the original version of this paper. This work was initiated
during the PIMS/APCTP/PITP Frontiers of Mathematical Physics Workshop on
``Particles, Fields and Strings'' at Simon Fraser University, Vancouver,
Canada, July 16--27 2001. The work of R.J.S. was supported in part by an
Advanced Fellowship from the Particle Physics and Astronomy Research
Council~(U.K.). L.D.P. would like to thank the MCTP for their hospitality
during his visit.

\end{document}